\documentclass{aa} 
\usepackage{graphicx}
\usepackage[flushleft]{threeparttable}
\usepackage[normalem]{ulem}
\usepackage{subcaption}
\usepackage{amssymb}
\usepackage{natbib}
\usepackage[colorlinks, citecolor=blue, linkcolor=blue, urlcolor=blue]{hyperref}
\usepackage{etoolbox}
\makeatletter
\patchcmd\@combinedblfloats{\box\@outputbox}{\unvbox\@outputbox}{}{%
   \errmessage{\noexpand\@combinedblfloats could not be patched}%
}%
 \makeatother

\begin{document}
\title{A New X-ray Selected Sample of Very Extended Galaxy Groups from the {\em{ROSAT}} All-Sky Survey}
\author{Weiwei Xu\inst{1,2,3}
\and
Miriam E. Ramos-Ceja\inst{2}
\and
Florian Pacaud\inst{2}
\and
Thomas H. Reiprich\inst{2}
\and
Thomas Erben\inst{2}\
}
\institute{National Astronomical Observatories, Chinese Academy of Sciences, Beijing, China\\
(\email{weiweixu@bao.ac.cn})
\and
Argelander-Insitut f{\"u}r Astronomie, Universit{\"a}t Bonn, Bonn, Germany
\and
University of Chinese Academy of Sciences, Beijing, China
}
\date{Accepted by A\&A}

\abstract
{ 
Some indications for tension have long been identified between cosmological constraints obtained from galaxy clusters and primary cosmic microwave background (CMB) measurements. Typically, assuming the matter density and fluctuations, as parameterized with ${\Omega}_\textrm{m}$ and ${\sigma}_{8}$, estimated from CMB measurements, many more clusters are expected than those actually observed. This has been reinforced recently by the Planck collaboration. One possible explanation could be that certain types of galaxy groups or clusters were missed in samples constructed in previous surveys, resulting in a higher incompleteness than estimated.
}{
    In this work, we aim to determine if a hypothetical class of very extended, low surface brightness, galaxy groups or clusters have been missed in previous X-ray cluster surveys based on the ROSAT All-Sky Survey (RASS).
}{
    We applied a dedicated source detection algorithm sensitive also to more unusual group or cluster surface brightness distributions. It includes a multiresolution filtering, a source detection algorithm, and a maximum likelihood fitting procedure. To optimize parameters, this algorithm is calibrated using extensive simulations before it is used to reanalyze the RASS data. In addition, the cross-correlation of the candidates with optical/infrared surveys is used for cluster identification and redshift estimation.
}{
    We found many known but also a number of new group candidates, which are not included in any previous X-ray or SZ cluster catalogs. In this paper, we present a pilot sample of $13$ very extended groups discovered in the RASS at positions where no X-ray source has been detected previously and with clear optical counterparts. The X-ray fluxes of at least $5$ of these are above the nominal flux-limits of previous RASS cluster catalogs ($\gtrsim 3 \times 10^{-12}$~erg~s$^{-1}$~cm$^{-2}$ in the $0.1-2.4$~keV energy band). They have low mass ($10^{13}~\rm{M}_{\odot} \lesssim$~$M_{500} \lesssim 10^{14}~\rm{M}_{\odot}$; i.e., they are galaxy groups), are at low redshift ($z<0.08$), and exhibit flatter surface brightness distributions than usual.
}{
    We demonstrate that galaxy groups were missed in previous RASS surveys, possibly due to the flat surface brightness distributions of this potential new population. Analysis of the full sample will show if this might have a significant effect on previous cosmological parameter constraints based on RASS cluster surveys.
}

\titlerunning{A sample of new extended galaxy groups from the RASS}
\authorrunning{W. Xu et al.}

\maketitle

\section{Introduction}
\label{sec:intr}

Groups and clusters of galaxies are powerful probes to test and constrain cosmological models, independent of and complementary to other methods (e.g., \citealt{Allen2011}), such as Type Ia supernovae \citep{Kowalski2008} and the cosmic microwave background (CMB, \citealt{Dunkley2009}). These systems have undergone gravitational relaxation and therefore are the largest objects whose masses can be measured with several independent methods. These measurements allow us to use the total matter content of clusters and cluster counts at different epochs to constrain the amount of dark matter in the Universe. Moreover, the rate of gravitationally driven structure formation as traced by the evolution of the cluster population provides independent constraints on the amount and properties of dark energy in the Universe.

Most cosmological studies that use galaxy groups and clusters require {\sl well-defined, large} cluster samples. To this purpose, cluster samples have to be carefully treated by understanding their underlying cluster populations as well as their cluster mass determination and calibration. This has become clear again in the recently identified {\sl tension} between primary CMB and cluster counts constraints on cosmological parameters encountered by the Planck collaboration (\citealt{Planck2014}, \citealt{Planck2016b}). Observational bias and the survey strategy must be carefully accounted for in order to provide a sufficiently accurate description of the observed cluster sample (e.g., \citealt{Mantz2010}, \citealt{Pacaud2016}, \citealt{Schellenberger2017}).

Early on, some tension was found between the values for $\sigma_8$, parameterizing the matter fluctuation amplitude, and $\Omega_\textrm{m}$, the mean normalized matter density. Cluster studies preferred $\sigma_8$ $\sim0.7$ (e.g., \citealt{Borgani2001}, \citealt{Reiprich2002}, \citealt{Seljak2002}, \citealt{Viana2002}) while CMB studies showed $\sigma_8$ $\sim0.9$ (e.g., \citealt{Spergel2003}) for a given $\Omega_\textrm{m}$ $\sim0.3$. After some oscillations (e.g., \citealt{Spergel2007}, \citealt{Henry2009}), now both probes seem to have settled around $\sigma_8$ $\sim0.8$ (e.g., \citealt{Mantz2010}, \citealt{Hinshaw2013}, \citealt{Haan2016}, \citealt{Planck2016b}, \citealt{Schellenberger2017}). Still, cluster constraints tend to be on the lower $\sigma_8$ side compared to primary CMB constraints while the statistical uncertainties shrink. This slight offset may simply reflect statistical uncertainty but could also be due to interesting systematic, physical, or cosmological effects (e.g., \citealt{Schellenberger2017}, \citealt{McCarthy2018}). Another probe at low-redshift, i.e.\ cosmic shear, also appears to result in lower $\sigma_8$ estimations (e.g., \citealt{Abbott2016}, \citealt{Hildebrandt2017}).

Among the observational techniques for the study of galaxy groups and clusters, X-ray imaging is one of the most sensitive and reliable methods to detect and analyze them. The presence of the hot intra-cluster gas ensures that only genuine gravitationally-bound structures are included, leading to highly pure and well-understood cluster samples in X-ray surveys.

The ROentgen SATellit (ROSAT) observatory has been the first and the only imaging telescope to perform an X-ray All-Sky Survey (RASS, \citealt{Truemper1992}, \citealt{Truemper1993}). The ROSAT mission was able to detect $\sim125\,000$ X-ray sources (\citealt{Voges1999}, \citealt{Voges2000}, \citealt{Boller2016}), providing an ideal basis for the construction of a large X-ray cluster sample for cosmological studies. In the northern hemisphere the largest and most representative galaxy cluster compilations from contiguous areas of RASS include the Bright Cluster Sample (BCS, \citealt{Ebeling1998}), which is a flux limited sample (flux $\geq4.4\times10^{-12}$~erg~s$^{-1}$~cm$^{-2}$ in the $0.1-2.4$~keV energy band), the extended BCS, which is a low flux extension of BCS (eBCS, flux $\geq2.8\times10^{-12}$~erg~s$^{-1}$~cm$^{-2}$ in the $0.1-2.4$~keV energy band, \citealt{Ebeling2000}), the Northern ROSAT All-Sky Survey (NORAS, \citealt{Boehringer2000}), which is not flux-limited and its selection is based on a minimum count-rate ($0.06$~cnts/s in the $0.1-2.4$~keV energy band, roughly corresponding to a flux $1.2\times10^{-12}$~erg~s$^{-1}$~cm$^{-2}$ in the same band, see their Figure~$8$a) and a source extent likelihood. The southern hemisphere has the ROSAT-ESO Flux Limited X-ray Galaxy Cluster Survey (REFLEX I - \citealt{Boehringer2001}), which is a flux limited sample (flux $\geq3\times10^{-12}$~erg~s$^{-1}$~cm$^{-2}$ in the $0.1-2.4$~keV energy band).

However, these galaxy cluster samples obtained from the RASS data could be incomplete. Because the detection method for most of such cluster samples, i.e.\ sliding cell algorithm (\citealt{Harnden1984}), works efficiently at finding point-like sources, instead of very extended sources with low surface brightness, such as galaxy groups and clusters at low redshift (e.g., \citealt{Valtchanov2001}). In addition, the inhomogeneous sky coverage of RASS might make the detection more difficult in regions with low or heavily varied exposure time. If the true incompleteness is higher than the estimated one, biased constraints on cosmological parameters may result. For the studies about cluster mass function, this would tend to lower the inferred values of $\Omega_\textrm{m}$ and/or $\sigma_8$, see, e.g., the tests carried out in \citet{Schellenberger2017}, shown as Figure~$10$ therein.

The main goal of this work is to explore possible missing extended galaxy groups or clusters and assess the completeness of the RASS-based cluster catalogs. 
For this purpose we reanalyze the RASS images\footnote{http://www.xray.mpe.mpg.de/cgi-bin/rosat/rosat-survey} in the $0.5-2.0$~keV energy band, with a wavelet-based source detection algorithm combined with a maximum likelihood fitting method, to detect and characterize X-ray sources. This procedure is chosen based on its successful performance in finding extended X-ray sources with low-surface brightness, i.e.\ galaxy groups and clusters at low redshift, in various X-ray surveys (e.g., \citealt{Rosati1995}, \citealt{Vikhlinin1998}, \citealt{Pacaud2006}, \citealt{Lloyd-Davies2011}). In this initial paper, we discuss the general X-ray characteristics of the possible missing groups and clusters in ROSAT catalogs, and show the properties of a pilot sample of confirmed detections. We refer to our extended detections as ``cluster candidates'' until we present the measured masses and we can determine if they are galaxy groups or clusters. Then, we call galaxy groups those with $M_{500}<10^{14}$~M$_\odot$ and galaxy clusters those with larger masses.

Throughout this paper, a Hubble constant $H_0=70~\rm km~s^{-1}~Mpc^{-1}$, matter density parameter $\Omega_\textrm{m}=0.3$, dark energy density parameter $\Omega_\Lambda=0.7$ are assumed if not stated otherwise. Moreover, the images used in this paper as well as the count-rates are in the $0.5-2.0$~keV energy band, while the fluxes are expressed in the $0.1-2.4$~keV energy band unless otherwise mentioned.

\begin{table*}[ht]
\begin{center}
\caption{\footnotesize{Summary of the extended source simulations in the ROSAT-like images. The column values indicate the number of simulated images for a given $\beta$-value.  For each set of parameters ($\beta$, $r_{\rm c}$ and flux within $r_{\rm c}$), $100$ clusters are simulated. For example, a value of 20 in the table means that there are 5 clusters in one simulated image for a given set of parameters. Simulations were performed for an exposure time of $450$~s with a background level of $0.08$~cnts/pixel, and four $\beta$-values ($0.40$, $0.55$, $0.66$ and $0.70$).}}
        \begin{tabular}{c | c c c c c c c c c c}
        \hline
        \hline
        Input flux ($0.5-2.0$ keV) & \multicolumn{10}{c}{Input core radius} \\
        $[$erg~s$^{-1}$~cm$^{-2}]$ & $0.75\arcmin$ & $1.50\arcmin$ & $2.25\arcmin$ & $3.00\arcmin$ & $3.75\arcmin$ & $4.50\arcmin$ & $5.25\arcmin$ & $6.00\arcmin$ & $12.00\arcmin$ & $24.00\arcmin$ \\
        \hline
    $1\times10^{-12}$ & $20$ & $20$ & $20$ & $20$ & $20$ & $20$ & $20$ & $20$ & $25$ & $100$\\ 
    $3\times10^{-12}$ & $20$ & $20$ & $20$ & $20$ & $20$ & $20$ & $20$ & $20$ & $25$ & $100$\\ 
    $5\times10^{-12}$ & $20$ & $20$ & $20$ & $20$ & $20$ & $20$ & $20$ & $20$ & $25$ & $100$\\ 
    $1\times10^{-11}$ & $20$ & $20$ & $20$ & $20$ & $20$ & $20$ & $20$ & $20$ & $25$ & $100$\\ 
    $5\times10^{-11}$ & $20$ & $20$ & $20$ & $20$ & $20$ & $20$ & $20$ & $20$ & $25$ & $100$\\ 
        \hline
        \hline
        \end{tabular}
\label{table:clu}
\end{center}
\end{table*}

\section{Cluster detection method}
\label{sec:method}

As stated in \citet{Rosati1995}, wavelet transformation is capable of detecting faint X-ray sources with a variety of sizes and surface brightness. \citet{Vikhlinin1998} has shown that a further implementation of a maximum likelihood method enhances the characterization of extended sources from the wavelet-detected sources. The combination of these methods has resulted in successful galaxy cluster catalogs using {\it XMM-Newton} data in recent years (e.g. \citealt{Pacaud2006}, \citealt{Lloyd-Davies2011}). Given that our main goal is to find possible missing extended groups and clusters with low surface brightness in the RASS data, we follow the general scheme and implement a detection pipeline similar to the one used in \citet[][hereafter P06]{Pacaud2006}. It proceeds in three main steps:
\begin{enumerate}
    \item The Poisson noise is first removed from the images with a multiresolution filtering procedure.
    \item The filtered images are scanned by a source detection algorithm, resulting in a preliminary source list.
    \item Each detected source is fitted and characterized by a maximum likelihood procedure.
\end{enumerate}
These steps are first applied to the simulations described in Section~\ref{sec:simu}, with the aim to assess the reliability of this method in finding extended clusters with low surface brightness; then this approach is applied to the RASS observations. In the following, a brief description of each step is presented.

The simulated images are first filtered using a wavelet algorithm designed to handle Poisson data. It is a completely new implementation of the wavelet task used in P06, \textsc{mr\_filter}\footnote{From the \textsc{mr/1} package: http://www.multiresolutions.com/mr/}. Like its predecessor, it relies on a ``$\grave{a}$ trous'' (``with holes'') transform with a cubic B-spline wavelet (also known as starlet transform) and filters the noise by imposing hard thresholds on the wavelet coefficients, estimated from autoconvolution histograms (see \citealt{StarckP1998} for a detailed description). A smoothed and denoised image is then reconstructed iteratively from the significant coefficients. A key feature of the new implementation is the ability to account for varying exposure time not only in the reconstruction, but also for the determination of the significance thresholds. More details on the software are provided in Faccioli, Pacaud et al. (submitted).

The source detection on the filtered images is performed by the \textsc{SExtractor}\footnote{http://www.astromatic.net/software/sextractor} software \citep{Bertin1996}. Originally, \textsc{SExtractor} was developed to detect objects from optical data, but it can also be applied to the multiresolution filtered X-ray images, since the filtering removes most of the noise from it to produce a smooth background. This step proceeds exactly as in P06 and the most relevant \textsc{SExtractor} parameters are given in their Table~1. The only difference is the pixel size, which we set to $45$~arcsec.

In the final step of the detection procedure, the sources identified by \textsc{SExtractor} are characterized by a maximum likelihood profile fitting algorithm (also a new implementation compared to P06). Starting from the \textsc{SExtractor} measurements as a guess for the source properties, it maximizes the Poisson likelihood for each source individually, masking surrounding detections based on the \textsc{SExtractor} segmentation map. The significance of each detection (hereafter {\it detection likelihood}) is then quantified using a relation based on the best fit likelihood as:
\begin{equation}
    \texttt{DET\_LIKE} = 2\,\left[\, \log{\left(\mathcal{L}_\mathrm{bgd}\right)} - \log{\left(\mathcal{L}_\mathrm{ext}\right)}
    \,\right],
\end{equation}
where ${\mathcal{L}_\mathrm{bgd}}$ and ${\mathcal{L}_\mathrm{ext}}$ are respectively the Poisson likelihoods of the best background (flat distribution) and extended source model. The extended source model is given by a $\beta$-profile, with $\beta$ value of $2/3$ (see Eq.~\ref{eq:beta}). A significance of the source extension (the {\it extension likelihood}) is also provided as:
\begin{equation}
    \texttt{EXT\_LIKE} = 2\,\left[\, \log{\left(\mathcal{L}_\mathrm{pnt}\right)} - \log{\left(\mathcal{L}_\mathrm{ext}\right)}
    \,\right],
\end{equation}
where ${\mathcal{L}_\mathrm{pnt}}$ is the likelihood of the point source model. It is worth mentioning that the point-like and the extended source models are convolved with the averaged ROSAT PSF. The source model parameter estimation is calculated through the likelihood ratio, which uses the C-statistic \citep{Cash1979} and it is minimized with the simplex method \textsc{AMOEBA} \citep{Press1992}.
Aside from the instrument specifics, the two main differences compared to P06 lie in the definition of the extension / detection likelihoods (which included an analytical correction for the number of fitted parameters in P06) and the background treatment. Indeed, our model includes a constant sky background to be fitted, whereas it was set in P06 through an analytical maximization of the likelihood with respect to the model normalization.

\section{Image simulations}
\label{sec:simu}

We first assess our cluster detection analysis on image simulations. For this, we perform extensive and dedicated Monte Carlo simulations of ROSAT images. Our procedure creates images with point-like and extended sources, which represent AGNs and galaxy clusters, respectively. It also takes into account the main ROSAT instrumental characteristics (such as point spread function, background and Poisson noise). In this section, we describe the setup of our simulations.

\subsection{General parameters}

We simulate images with the same size as the publicly available RASS tiles, i.e.\ $6.40\times6.40$~deg$^2$, with a pixel size of $45$~arcsec. 

The input sources are convolved with the survey averaged point spread function (PSF) at $1.0$~keV (as appropriate for the $0.5-2.0$~keV energy band that we are simulating).
The PSF itself is obtained from the Gaussian model of \cite{Boese2000}. This model describes the width of the PSF (in units of arcsec) as a function of energy and off-axis:
\begin{equation}
    \sigma(E,\epsilon) = \sqrt{108.7E^{-0.888}+1.121E^6 +0.219\epsilon^{2.848}}
\end{equation}
where $E$ is the energy expressed in keV and $\epsilon$ is the off-axis angle in arcmin. For the field averaged PSF, applicable to survey observation, we generate Gaussian PSF at different off-axis angles and weight them according to the telescope vignetting and detector area. The resulting PSF has a full-width-half-maximum (FWHM) of $\sim~45$~arcsec and, despite originating from a sum of Gaussian functions, has much broader wings than a single Gaussian. 

The {\em{ROSAT}} survey exposure distribution is not uniform. It varies from $100$~s to $40\,000$~s at the ecliptic equator and poles, respectively \citep{Voges1999}. Given that the purpose of this work is to characterize the average potential of our extended source detection algorithm and to provide insights on the properties of new clusters, we adopt a flat exposure time of $450$~s for the simulations. This value is the average obtained over all exposure maps of the ROSAT all-sky survey. In a following paper, the limitations of this approach will be fully discussed.

We also include a flat X-ray background with a value of $0.08$~cnts/pixel. This value is calculated by averaging the background maps of the ROSAT all-sky survey. These maps are created by the \textsc{SExtractor} software (see Section~\ref{sec:method} for further details).

\begin{figure*}[ht]
\centering
\includegraphics[width=1.0\textwidth]{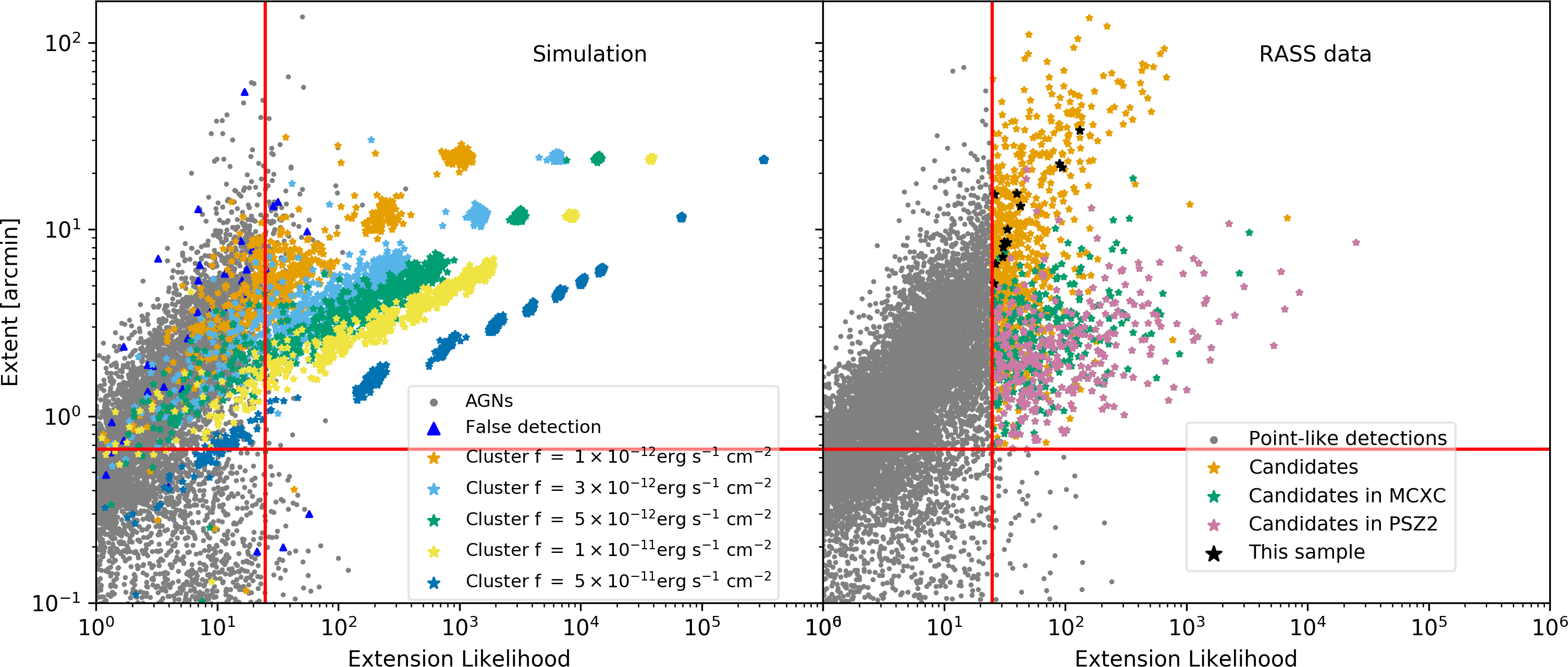}
\caption{\footnotesize
{Selection criteria for extended sources. The selection is performed in the extension likelihood - extent plane. The red-solid lines define the optimal parameters obtained from simulations to characterize extended sources. {\sl Left:} Simulation results. Grey dots represent simulated AGNs, and blue triangles, false detections. Coloured stars represent clusters with different input fluxes. {\sl Right:} Results from reprocessing RASS data. Grey dots stand for point-like detections and the star symbols for the cluster candidates. Green and pink stars show the candidates with identified counterparts in the MCXC and PSZ2 catalogs, respectively. The black stars represent the $13$ groups of our pilot sample described in Table~\ref{tab:candi}.}}
\label{fig:extextlikeplane}
\end{figure*}

\subsection{Galaxy clusters and AGNs}

The surface brightness, $S_\textrm{X}$, of galaxy clusters is described by the spherically symmetric $\beta$-model \citep{Cavaliere1976}. It is given by
 \begin{equation}
S_\textrm{X}(r)~\propto \left[{1~+~\left(\frac{r}{r_{\rm c}}\right)^2}\right]^{-3{\beta}~+~1/2},
\label{eq:beta}
 \end{equation} 
where $r_\textrm{c}$ is the core radius of the cluster, and $\beta$ determines the slope of the brightness profile. Different values of $\beta$, $r_\textrm{c}$ and flux (see Table~\ref{table:clu}) are used in order to cover a wide range of possible cluster morphologies. For each set of parameters, $100$ clusters are simulated with controlled positions. In order to reduce the overlap of X-ray emission from the simulated clusters, the distances between them are larger than $2\times r_{500}$. Moreover, the cluster distances to the nearest image edge are always larger than $r_{500}$. The value of $r_{500}$ is taken as $7\times r_{\rm c}$.

Since AGNs represent the majority of extragalactic X-ray sources, a population of them are included in the image simulations. The surface brightness of AGNs is described by the {\em{ROSAT}} PSF. The AGN flux distribution and source density are obtained using the $\log N-\log S$ from \citet{Moretti2003} down to a flux of $2.29\times10^{-14}$~erg~s$^{-1}$~cm$^{-2}$. The spatial distribution of AGNs is assumed to be random. 

To convert the flux of cluster and AGN into photon count, PIMMS\footnote{https://heasarc.gsfc.nasa.gov/cgi-bin/Tools/w3pimms/w3pimms.pl} is used. The energy conversion factor used for clusters is $1.12\times10^{-11}$~(erg~s$^{-1}$~cm$^{-2}$)/(cnts~s$^{-1}$), which is obtained by assuming an absorbed APEC model with $n_{\rm{H}}=5.95\times 10^{20}$~cm$^{-2}$, an abundance of $0.40$ times of the solar abundance, a temperature of $2.73$~keV ($10^{7.5}$~K) and $z=0.10$. The galactic neutral hydrogen column density value, $n_{\rm{H}}$, is taken from the average of all observations, excluding only a $\pm20$~deg area around the Galactic plane. The values of the temperature and redshift are chosen based on our objective of detecting relatively low-temperature and nearby clusters previously missed. Note that the energy conversion factor changes by less than $3\%$ for $0.02<z<0.18$ and $T=2.73$~keV. The energy conversion factor for AGNs is $1.15\times10^{-11}$~(erg~s$^{-1}$~cm$^{-2}$)/(cnts~s$^{-1}$). We obtain this factor by assuming an absorbed power-law model with photon index equal to $1.85$, which is roughly the peak of the observed distribution for X-ray AGNs \citep{Ueda2014}, and with the same $n_{\rm{H}}$ value as before.

\subsection{Source classification using simulations}
\label{subsec:class_simu}

All simulated images were processed with the pipeline described in Section~\ref{sec:method}. The positions of the detected sources were cross-identified with the simulation inputs using a correlation radius of $6$~arcmin for point-like sources and $3.5\times r_{\textrm{c}}$ for extended sources. We use a variable matching radius for extended sources because their positional accuracy is size dependent.

Following P06, we explore the parameter space of the maximum likelihood fitting method to place source classification criteria. The most relevant parameters are: detection likelihood, extension likelihood, source extent (i.e.\ core radius) and source position.

The wavelet image reconstruction creates artifacts near the image edges, which result in false detections. Therefore, we exclude all detected sources located close to the image edges ($<15$ pixels to image sides). Moreover, our analysis shows that a threshold value of $20$ in the detection likelihood removes $95.4\%$ of the false detections, i.e.\ sources that are not simulated but are found by the detection algorithm. Although this threshold removes $50.8\%$ of the detected AGNs, it preserves $96.7\%$ of the detected clusters. We obtain an average of $\sim1.5$ false detections per simulated image. The number of $1.5$ false detections is a global value of the number of false detections regardless of the origin of the simulated source (point-like or extended). However, as it is discussed in the following, this number is mainly generated by false point-like detections.

\begin{figure}[t]
\centering
\includegraphics[width=1.0\columnwidth]{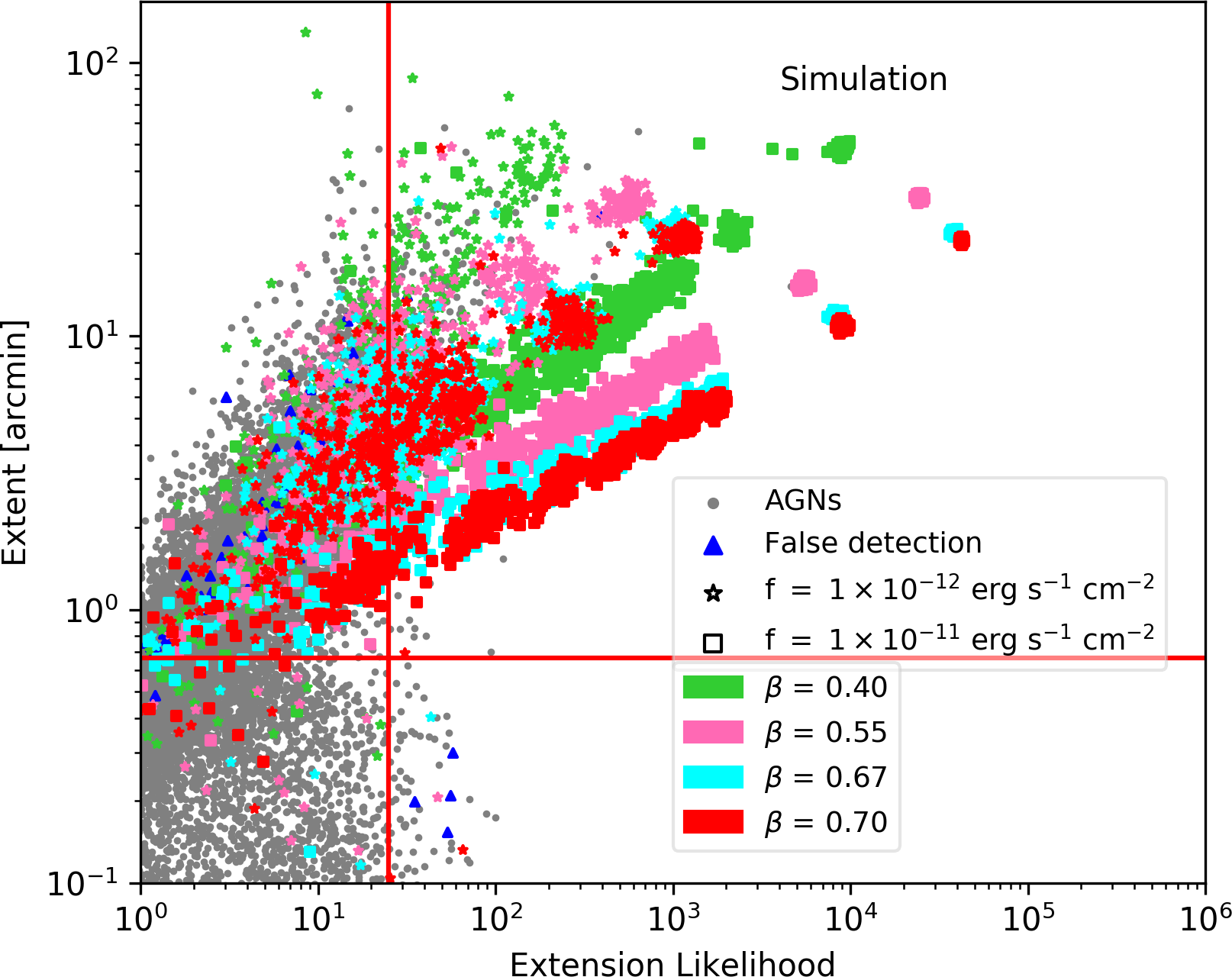}
\caption{\footnotesize{Similar plot as the left panel of Figure~\ref{fig:extextlikeplane} showing the selection criteria for extended sources in the extension likelihood - extent plane. Grey dots represent simulated AGNs, and blue triangles, false detections. Coloured symbols represent clusters with different input $\beta$-values. Filled stars show clusters with input flux $1\times10^{-12}$~erg~s$^{-1}$~cm$^{-2}$, while filled squares are clusters with input flux $1\times10^{-11}$~erg~s$^{-1}$~cm$^{-2}$.}}
\label{fig:diffbeta_Sel}
\end{figure}

By scanning the extension likelihood - extent parameter space, we look for a location where the majority of our simulated extended sources are recovered while maintaining the contamination from point-like sources at a low level. The left panel of Figure~\ref{fig:extextlikeplane} shows the main selection criteria for extended sources with $\beta=2/3$ in this plane. This is given by values of extension likelihood $>25$ and extent $>0.67$~arcmin. The discrete distribution in the extent values reflects the different input core radius values used in the simulations (see Table.~\ref{table:clu}). Using these extended selection criteria we obtain $\sim 4.2\times10^{-3}$ extended false detections per simulated image, and $\sim0.1$ misclassified AGNs per simulated image, i.e. simulated AGN with values of extension likelihood $>25$ and extent $>0.67$~arcmin.

The minimum extent value chosen for extended source characterization, $0.67$~arcmin, is slightly lower than the minimum value of core radius used in the simulations ($0.75$~arcmin). We have tested the impact of this minimum extent value, by simulating clusters with $r_\textrm{c}=0.5$~arcmin. We found that these clusters are characterized by our algorithm with low values of extension likelihood, $<25$. Therefore, they are not characterized as extended sources. This does not represent an issue in the present work since our goal is to look for very extended galaxy clusters that might have been missed in previous studies.

Figure~\ref{fig:diffbeta_Sel} shows that the selection criteria are very similar for extended sources simulated with $\beta$-values different from $2/3$. In contrast with the left panel of Figure~\ref{fig:extextlikeplane}, this plot displays only two values of input cluster fluxes, $1\times10^{-12}$~erg~s$^{-1}$~cm$^{-2}$ and $1\times10^{-11}$~erg~s$^{-1}$~cm$^{-2}$. In general, clusters simulated with small $\beta$-values tend to have lower extension-likelihood values and larger extent values.

The detection and characterization efficiency of the total number of simulated clusters can be summarized as following. Approximately $34\%$ of simulated clusters with $\beta=0.40$, flux equal to $1\times10^{-12}$~erg~s$^{-1}$~cm$^{-2}$ and $r_\textrm{c}>5'$ are detected and characterized as extended sources. This percentage increases to $\sim97\%$ for clusters with flux equal to $3\times10^{-12}$~erg~s$^{-1}$~cm$^{-2}$. About $69\%$ of simulated clusters with $\beta=0.55$, flux equal to $1\times10^{-12}$~erg~s$^{-1}$~cm$^{-2}$ and $r_\textrm{c}>5'$ are detected and characterized as extended sources. Broadly speaking, simulated clusters with fluxes $\ge 5\times10^{-12}$~erg~s$^{-1}$~cm$^{-2}$, $\beta \ge 0.55$ and $r_\textrm{c}\ge 5'$ have $100\%$ probability of being detected and characterized as extended sources. The full analysis about the completeness of our sample will be assessed in a subsequent publication.

\begin{figure}[t]
\centering
\includegraphics[width=1.0\columnwidth]{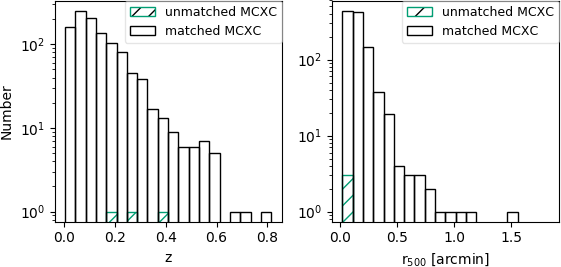}
\caption{\footnotesize{Distribution of redshift and $r_{\rm 500}$ of matched and unmatched RASS-based MCXC clusters with our detected sources. Notice the log-scale on the y-axis.}}
\label{fig:x_hist_mcpl_pp.png}
\end{figure}

\section{Results}
\label{sec:obs}

\subsection{Performance on real data}
\label{subsec:class_obs}

Having calibrated the cluster candidate selection criteria on simulations, we proceed to run the detection algorithm on the real RASS data. The RASS data consist of $1\,378$ image tiles, with varying exposure time (see Section~\ref{sec:simu}). We exclude the data close to the plane of the Milky Way, the Magellanic Clouds and the Virgo cluster regions (see Table~2 of \citealt{Reiprich2002}).

The right panel of Figure~\ref{fig:extextlikeplane} shows the results of processing the RASS data. Applying the same criteria for extended sources as in Section~\ref{subsec:class_simu}, we find $1\,308$ cluster candidates. In order to validate the authenticity of these cluster candidates, we first cross-identify them with well-known cluster catalogs.

Using a $10$~arcmin correlation radius, we match our cluster candidates with the Meta-Catalog of X-Ray Detected Clusters of Galaxies (MCXC, \citealt{Piffaretti2011}) and the second {\em{Planck}} catalogue of Sunyaev-Zeldovich sources (PSZ2, \citealt{Planck2016a}), excluding clusters in the regions where we do not process the RASS data. The size of the correlation radius is determined from the distribution of the positional difference between our cluster candidates and the MCXC and PSZ2 clusters. Within $10$~arcmin the differential number of matched clusters decreases rapidly. Beyond this value, the positional difference distribution starts to raise, meaning that random matches start to appear due to the increasing area. 

First, we correlate all our detections (i.e. regardless of their extent, detection likelihood and extension likelihood values), with clusters of the MCXC and PSZ2 catalogs. We are able to detect $86.5\%$ of the MCXC clusters. This percentage increases to $99.7\%$ if we only select RASS-based MCXC clusters. Figure~\ref{fig:x_hist_mcpl_pp.png} shows the redshift and $r_{500}$ (taken from \citealt{Piffaretti2011}) distributions of the matched and unmatched RASS-based MCXC clusters with our detected sources. From Figure~\ref{fig:x_hist_mcpl_pp.png} we can see that we do not detect $3$ RASS-based MCXC clusters. Two of such clusters are located in regions with exposure time $<70$~s, while the third one is positioned close to the ROSAT North Ecliptic Pole where the large gradient of exposure time may affect the positional accuracy of the detected sources in the region (our nearest detection is $12$~arcmin away). We also observe a high total detection rate of $97\%$ of PSZ2 clusters with $z<0.2$. This percentage decreases to $93\%$ when correlating our detections with all PSZ2 clusters.

\begin{figure}[t]
\centering
\includegraphics[width=1.0\columnwidth]{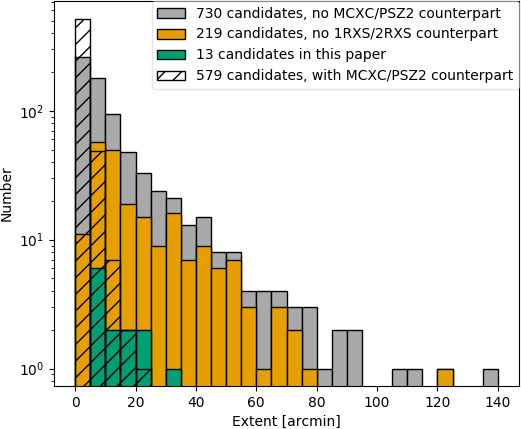}
\caption{\footnotesize{Extent value distribution for the $1\,308$ cluster candidates. Out of these $1\,308$ candidates, $578$ have a MCXC or PSZ2 counterpart (filled pattern histogram) and $730$ do not have a counterpart in these catalogs (grey histogram). Out of the $730$ candidates, $219$ do not possess a counterpart in the 1RXS or 2RXS catalog (orange histogram). The green histogram shows the extent distribution of the $13$ group candidates discussed in detail in this work (see Section~\ref{sec:pilotsample}). Notice the log-scale on the y-axis.}}
\label{fig:histononmatch}
\end{figure}

Using the whole MCXC and PSZ2 catalogs and imposing our extended source criteria, out of the $1\,308$ cluster candidates, $540$ have a counterpart in MCXC, while $323$ have a correspondence in PSZ2 (Figure~\ref{fig:histononmatch}). These numbers overlap in $285$ clusters, therefore we remain with $730$ cluster candidates without counterparts in MCXC or PSZ2. The extent distribution of the cluster candidates is displayed in Figure~\ref{fig:histononmatch}. We find that $\sim67\%$ of our cluster candidates with extent $\lesssim 5$~arcmin have a counterpart in the MCXC or PSZ2 catalogs, showing the reliability of our extended source finder. 
The grey histogram in Figure~\ref{fig:histononmatch} shows the distribution of the cluster candidates with no counterpart in MCXC or PSZ2. Approximately $60\%$ of these candidates have an extent $\gtrsim 5$~arcmin, leaving very extended sources as possible candidates for missed clusters in the RASS-based cluster catalogs.

\subsection{Looking for new galaxy clusters and groups}
\label{sec:newclu}

As mentioned previously, the purpose of this paper is not to describe a new cluster catalog obtained by reprocessing the RASS data, but to show a pilot sample of promising cluster and group discoveries.

We further investigate the cluster candidates that have not been previously detected in the MCXC or PSZ2 catalogs. We use a correlation radius of $5$~arcmin to match our $730$ cluster candidates with the the First {\em{ROSAT}} all-sky survey source catalogue (1RXS, which includes the {\em{ROSAT}} All-Sky Survey Bright Source Catalogue - BSC - \citealt{Voges1999} and the {\em{ROSAT}} All-Sky Survey Faint Source Catalogue - FSC - \citealt{Voges2000}), and the Second {\em{ROSAT}} all-sky survey source catalogue (2RXS, \citealt{Boller2016}). We find $219$ candidates with no detections in 1RXS or 2RXS. The distribution of these cluster candidates is shown in the orange histogram of Figure~\ref{fig:histononmatch}, which shows larger values of extent than the candidates matched in the 1RXS or 2RXS catalogs. We concentrate from now on only on these cluster candidates.

To further validate the $219$ cluster candidates we perform an exhaustive visual inspection and literature search. For the visual inspection, we used the ROSAT photon images, and their counterparts in infrared, optical and microwave bands (2MASS\footnote{https://www.ipac.caltech.edu/2mass}, SDSS\footnote{http://www.sdss.org}, DSS\footnote{http://archive.eso.org/dss/dss}, ATLAS\footnote{http://astro.dur.ac.uk/Cosmology/vstatlas} and {\em{Planck}}\footnote{http://www.esa.int/Our$\_$Activities/Space$\_$Science/Planck}) overlaid with X-ray contours obtained from the wavelet-filtered images. The ROSAT exposure maps are also checked to make sure that the most extended detections are not artificially created by large exposure gradients. We also use images of the HI~$4\pi$ survey (HI4PI, \citealt{HI4PI2016}) to assure that the cluster candidates are also not due to inhomogeneities in the HI sky distribution. We checked if our cluster candidates have already been discovered by means of other X-ray, optical or SZ surveys, and whether our source candidates are nearby galaxies given their extension. Table~\ref{table:catalog} shows the catalogs used to cross-check our candidates. Furthermore, we include spectroscopic redshift distributions obtained from SDSS~DR14\footnote{http://www.sdss.org/dr14/} or the NASA Extragalactic Database\footnote{https://ned.ipac.caltech.edu/} (NED) to assess the reliability of the cluster candidates. We do not restrict the redshift values when constructing these distributions.

Using the above information we subjectively rated each of the $219$ cluster candidates with different grades: $1$ for cluster candidates that are already known from the cluster catalogs shown in Table~\ref{table:catalog}, $2$ for cluster candidates that show a clear galaxy over-density in the infrared and/or optical images, $3$ for cluster candidates for which is not clear that a galaxy over-density exists due to lack of optical and/or infrared data, and $4$ for cluster candidates that we consider as noise, e.g. X-ray detections created by strong gradients of exposure in the RASS data or by low density value of HI in small regions of the sky. Our final grades show that $20\%$ of the $219$ cluster candidates are rated with $1$, $16\%$ with $2$, $20\%$ with $3$ and $44\%$ with $4$. The number of cluster candidates rated as noise corresponds to $0.1$~false extended detections per image. This rate is consistent with the number of misclassified AGNs and false extended sources per image in our simulations (see Section~\ref{subsec:class_simu}).

\begin{table}[t]
\caption{\footnotesize{Overview of the publications used to cross-check if cluster candidates have been previously detected.}}
    \begin{threeparttable}
    \centering
        \begin{tabular}{p{0.4\linewidth}
                        p{0.5\linewidth}}
        \hline
        \hline
            X-ray catalogs &\citealt{Ebeling1996}\\
                           & \citealt{Mehrtens2012} \\
                           &  \citealt{Liu2015}  \\
                           &  \citealt{Pacaud2016} \\
                           &  \citealt{Wen2018}  \\
        \hline
            Optical catalogs &\citealt{Zwicky1968}  \\
                             &\citealt{Abell1989}  \\
                             &\citealt{Wen2009} \\ 
                             &\citealt{Wen2012} \\ 
                             &\citealt{Rykoff2014}\\
                             & \citealt{Rykoff2016} \\
                             & \citealt{Oguri2018} \\
        \hline
           Microwave catalogs & \citealt{Marriage2011} \\
                             & \citealt{Hasselfield2013}  \\
                             &  \citealt{Hilton2017} \\
        \hline
          Galaxy catalogs    & \citealt{Zwicky1968}\\
                             & \citealt{Sinnott1988} \\ 
                                 
        \hline
        \hline
        \end{tabular}
        \end{threeparttable}
\label{table:catalog}
\end{table}

In this paper, we present a detailed analysis of a pilot sample of $13$ cluster candidates selected out of those rated as $2$. In optical, these cluster candidates coincide with clear over-densities of bright elliptical galaxies (see Figure~\ref{fig:clustercandidates}) distributed over extended areas ($\gtrsim30$~arcmin) and NED shows that $11$ out of the $13$ cluster candidates have been previously reported in optical group and cluster catalogs. Furthermore, for all the candidates we are able to determine a reliable redshift estimation from existing galaxy catalogs (see Section~\ref{sec:redshift}). In the following section, we describe the methods used to characterize each individual cluster.

\subsection{Pilot sample of cluster candidates }
\label{sec:pilotsample}

In this section, we describe the methods used to characterize the properties of $13$ cluster candidates. Table~\ref{tab:candi} shows the sky coordinates of each system. The locations of these candidates in the extension likelihood - extent plane are shown with the black star symbols in the right panel of Figure~\ref{fig:extextlikeplane}.

\subsubsection{Redshift estimation}
\label{sec:redshift}

We use NED to determine the redshift of each cluster candidate. We select all galaxies with spectroscopic redshift and within a box of size $1\times1$~deg$^2$ centered on the cluster candidate position. For most of the cluster candidates, there are none or very few galaxies with redshift above $0.1$. The final redshift depends on the number of available redshifts and their distribution. We fit iteratively a Gaussian function to the redshift distribution in $z<0.1$ range, and apply $3\sigma$ clipping to remove outliers. The redshift of candidate is derived from its peak (see right panels Figure~\ref{fig:clustercandidates}). For the cases with two peaks in the redshift distribution, we choose the closest redshift peak to the redshifts of the optical groups/clusters found in NED. Table~\ref{tab:candi} displays the total number of galaxies used for the redshift determination and the right panels of Figure~\ref{fig:clustercandidates} shows their redshift distribution for each cluster candidate. We find that our $13$ cluster candidates are in the redshift range $0.013-0.072$, with a median value of $0.036$. $11$ of them are below $z<0.05$.

According to NED, $11$ out of our $13$ group candidates has some published counterparts within $15$~arcmin (see Section~\ref{sect:notes} for details). We look into these catalogs to corroborate our redshift estimation. The red lines in the right panels of Figure~\ref{fig:clustercandidates} represent the redshifts of these matched groups or clusters showing a good agreement with our estimation.

\subsubsection{Flux, luminosity and mass determination}
\label{sec:flux_estim}

Finally, we ran a more involved photometric analysis using the growth curve analysis method (hereafter GCA, \citealt{Boehringer2000, Boehringer2001}), which computes the ROSAT hard band, $0.5-2$~keV, integrated count-rate as a function of aperture around the estimated cluster center. For this, a local background is first estimated in a large annulus, then the background subtracted count-rate is estimated at lower radii in bins of $0.5^{\prime}$ and summed within increasing apertures. Contamination by surrounding sources is corrected for by a deblending procedure that separates the signal in a number of angular sectors and excludes those falling outside $2.3\sigma$ above/below the median value. The error on aperture count-rates rely on Poisson statistics and is combined with the uncertainty on the background count rate. By default, the GCA method over ROSAT data uses a background annulus of $20-41.3^{\prime}$, but given the large size of our cluster candidates, we had to increase these values manually for most sources, reaching up to $85-110^{\prime}$ for the most extended source.

At this point, we obtain a full radial profile of the source and a decision is needed as to the aperture within which the source flux should be measured. We follow closely the methods and assumptions laid out in \citet{Boehringer2013} for the REFLEX-II cluster sample so that our flux measurements are comparable to this reference catalogue. At first, a significance radius, $r_\textrm{sig}$, is determined for which the 1$\sigma$ uncertainty interval encompasses all count-rates integrated in larger apertures. Then, the count-rate within this aperture, $CR_\textrm{sig}$, is estimated by fitting a constant value to the plateau outside $r_\textrm{sig}$. Since this aperture holds no physical meaning for the general cluster population, we convert it to a more meaningful value, within the fixed spherical overdensity $r_{500}$ based on a few assumptions on the appearance of galaxy clusters.

For a given galaxy cluster mass, $M_{500}$, the integrated $0.1-2.4$~keV luminosity, $L_{500}$, and average spectroscopic temperature, $T_\textrm{X}$, are estimated from the \citet{Reichert2011} scaling relations. The corresponding flux and count-rate within $r_{500}$, which we refer to as $F_{500}$ and $CR_{500}$, are derived from the ROSAT PSPC response using appropriate corrections for spectral redshifting, flux dilution and absorption. Then a $\beta$-model with $\beta=2/3$ and core radius $r_\textrm{c}=r_{500}/7$ is assumed to convert $CR_{500}$ to an expected count-rate in the significance radius. This mapping enables us to iterate over cluster mass until the right count-rate is obtained within $r_\textrm{sig}$. This provides us with a set of estimated cluster parameters derived from the model ($r_{500}$, $F_{500}$, $T_\textrm{X}$, $L_{500}$ and $M_{500}$).

\begin{figure}[t]
\centering
\includegraphics[width=1.0\columnwidth]{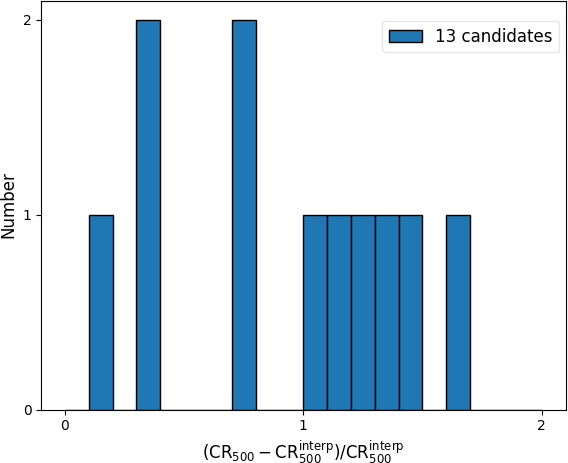}
\caption{\footnotesize{Distribution of the normalized difference between $CR_{500}$ and $CR_{500}^\textrm{interp}$ (see Section~\ref{sec:flux_estim} for details on the estimation of these quantities).}}
\label{fig:x_hist_cr_ratio_trim.png}
\end{figure}

The third panel of each row in Figure~\ref{fig:clustercandidates} show the growth curves estimated following the REFLEX-II procedure described above. As conspicuous from these plots, the observed growth curves (solid lines) are much flatter than the surface brightness distribution predicted by the assumed $\beta$-model (dotted lines). This confirms that our detection algorithm permits to identify very extended low surface brightness sources, that may be missing from previous catalogs. A downside, however, is that the assumed cluster profiles are not suitable to reliably convert fluxes from one radius to the other.
Fortunately, the significance radius of the cluster emission exceeds the estimated value of $r_{500}$ for all of our detections.
This permits to by-pass the use of a fixed surface brightness model by comparing directly the count-rate  predicted by our scaling relation model for a given $M_{500}$ with the observed value within the corresponding $r_{500}$. A revised estimate of the count-rate, $CR_{500}^{\mathrm{interp}}$, is thus obtained by interpolating over the growth-curve such that $CR(r_{500})/CR_{500}(M_{500})=1$.
We note, however, that this new estimate of $CR_{500}^\textrm{interp}$ still includes a correction based on the default $\beta$-model, since we use it to estimate the dilution resulting from the ROSAT PSF -- but it only makes a tiny difference for such extended systems. In addition, given the peculiar properties of our sources, they may not follow the average scaling relations used in the model. Consequently, although our quoted masses and fluxes always rely on the interpolated measurements, we preferred to list in Table~\ref{tab:candi2} both values of $CR_{500}$ and $CR_{500}^\textrm{interp}$ as an illustration of systematic uncertainties.
The differences between the two estimates are strongly variable (see Figure~\ref{fig:x_hist_cr_ratio_trim.png}), but the values of $CR_{500}$ are always overestimated by a factor of $2.05$ on average.
 
The main characteristics of this pilot sample can be summarized as following. The $13$~cluster candidates have $r_{500}\gtrsim8$~arcmin, reaching values up to $\sim30$~arcmin. $5$ of the $13$ systems have flux measurements larger than the flux limit values of previous RASS cluster catalogs ($\gtrsim 3\times10^{-12}$~erg~s$^{-1}$~cm$^{-2}$, see Section~\ref{sec:intr}).
All $13$ cluster candidates have $M_{500}\lesssim10^{14}~$M$_\odot$, placing them in the group regime, rather than in the cluster scale. Therefore, from now on we will refer to our $13$ cluster candidates as galaxy groups.

\subsubsection{Dynamical mass estimation}
\label{sec:m_sigmav}

\begin{figure}[t]
\centering
\includegraphics[width=1.0\columnwidth]{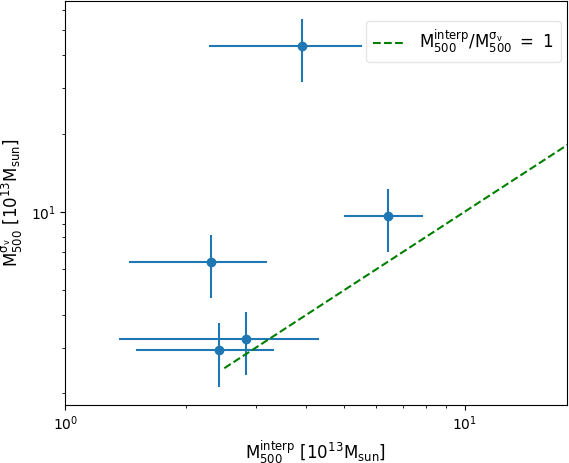}
\caption{\footnotesize{Comparison of the derived dynamical masses with the masses derived from the $L_\textrm{X}-M$ relation (see Section~\ref{sec:m_sigmav} for details).}}
\label{fig:m_m}
\end{figure}

For the galaxy groups with more than $20$ galaxy members with spectroscopic information (groups no.~$3,~8,~9,~10,~12$), we also calculated their dynamical masses from their velocity dispersion. First, the velocity dispersion is obtained from the fit of the Gaussian function with a $3\sigma$ clipping to the redshift distribution. Then, using the Eq.~1 from \citet{Munari2013}, which provides the relation between velocity dispersion and $M_{200}$, a range estimation of $M_{200}^{\sigma_{v}}$ is derived. We further assume a Navarro-Frenk~and~White (NFW, \citealt{Navarro1997}) profile with concentration parameter $c=r_{200}/r_\textrm{s}=4$ to determine a range of $M_{500}^{\sigma_{v}}$ by using $r_{500} \approx 0.65~r_{200}$ \citep{Reiprich2013}. The last column of Table~\ref{tab:candi2} shows the range of values of the dynamical mass estimations. Figure~\ref{fig:m_m} shows that the dynamical masses are in general agreement with the masses derived from the $L_\textrm{X}-M$ relation (see Section~\ref{sec:flux_estim}). Only group no.~$8$ shows a significant difference between these two mass estimations, this is further discussed in Section~\ref{sec:clu8}.

\begin{table}[t]
  \begin{threeparttable}
  \caption{\footnotesize{Galaxy groups in this pilot study. The sample is sorted according to the extent value of each group. Then, each group is identified using a sequential number (column 1) to simplify their discussion in this paper. $N_\textrm{g}$ is the number of galaxies used to estimate the redshift of candidates.}}
  \label{tab:candi}
  \centering
  \begin{tabular}{c c c c  c}
  \hline
  \hline
  Group & R.A. & Dec.  & $z$ & $N_\textrm{g}$\\
    No.  & J2000  & J2000  &   &  \\
\hline
$~1$&$21~56~59.76$ &$+28~00~57.6~$ &$0.069$&$~~6$\\
$~2$&$03~04~02.64$ &$-12~04~22.8~$ &$0.013$&$~10$\\
$~3$&$10~50~21.60$ &$+00~16~44.4~$ &$0.039$&$~42$\\
$~4$&$01~05~19.44$ &$+36~00~50.4~$ &$0.072$&$~~7$\\
$~5$&$02~25~57.12$ &$+36~59~45.6~$ &$0.036$&$~17$\\
$~6$&$01~36~51.36$ &$-14~00~25.2~$ &$0.040$&$~14$\\
$~7$&$23~17~25.68$ &$+29~02~31.2~$ &$0.023$&$~~8$\\
$~8$&$16~06~24.00$ &$+15~40~12.0~$ &$0.040$&$123$\\
$~9$&$16~50~22.32$ &$+23~35~42.0~$ &$0.036$&$~52$\\
$10$&$13~49~18.96$ &$-07~13~48.0~$ &$0.025$&$~22$\\
$11$&$21~07~18.24$ &$-47~10~08.4~$ &$0.017$&$~10$\\
$12$&$13~28~56.88$ &$-28~02~27.6~$ &$0.033$&$~38$\\ 
$13$&$02~41~01.92$ &$+08~46~55.2~$ &$0.021$&$~14$\\
  \hline
  \hline
  \end{tabular}
  \end{threeparttable}
\end{table}

\begin{sidewaystable*}
 \begin{center}
  \caption{\footnotesize{Photometry and mass determination for the new galaxy group candidates, refer to Section \ref{sec:pilotsample} for details.}
   }
   \label{tab:candi2}
    \begin{threeparttable}
    \centering
        \begin{tabular}{c c c c c c c c c c c }
\hline
\hline
Group &   Extent\ \tnote{+} & $r_\textrm{sig}$\ \tnote{$\ddagger$} & $CR_\textrm{sig}$\ \tnote{$\ddagger$} & $r_\textrm{500}$\ \tnote{$\ast$} & $CR_\textrm{500}$\ \tnote{$\ast$} & $CR_\textrm{500}^\textrm{interp}$\ \tnote{$\dagger$} & $F_\textrm{500, 0.1-2.4~keV}^\textrm{interp}$\ \tnote{$\dagger$} & $L_\textrm{500, 0.1-2.4~keV}^\textrm{interp}$\ \tnote{$\dagger$} & $M_\textrm{500}^\textrm{interp}$\ \tnote{$\dagger$} & $M_\textrm{500}^{\rm \sigma_v}$\ \tnote{$\star$}  \\
      No.    &  [arcmin]  &[arcmin]&   [cnts/s]    &[arcmin]   &[cnts/s]     &[cnts/s]                     & [$10^{-12}$~${\rm erg~s^{-1}~cm^{-2}}$]& [$10^{42}$~${\rm erg~s^{-1}}$]& [$10^{13}$~M$_{\odot}$]& [$10^{13}$~M$_{\odot}$] \\
      \hline
      $~1$&$~5.11$&$11.00$ & $0.06\pm0.02 $ & $~7.90\pm0.53 $ & $0.06 $ & $0.05\pm0.01$ & $0.97\pm0.31$ & $11.09\pm3.53$ & $6.14\pm1.23$ & -- \\             
      $~2$&$~6.56$&$28.50$ & $0.45\pm0.12 $ & $28.86\pm1.91 $ & $0.45 $ & $0.46\pm0.12$ & $7.74\pm2.37$ & $~2.82\pm0.86$ & $2.70\pm0.51$ & -- \\             
      $~3$&$~7.10$&$19.00$ & $0.16\pm0.04 $ & $12.60\pm0.75 $ & $0.15 $ & $0.06\pm0.02$ & $0.89\pm0.37$ & $~3.12\pm1.31$ & $2.84\pm0.74$ & [$2.79-3.67$]\\   
      $~4$&$~8.05$&$27.00$ & $0.20\pm0.04 $ & $~9.72\pm0.44 $ & $0.18 $ & $0.07\pm0.02$ & $1.52\pm0.36$ & $18.25\pm4.22$ & $8.36\pm1.20$ & -- \\             
      $~5$&$~8.53$&$19.00$ & $0.22\pm0.05 $ & $14.30\pm0.67 $ & $0.21 $ & $0.19\pm0.03$ & $3.84\pm0.71$ & $11.39\pm2.10$ & $6.34\pm0.73$ & -- \\             
      $~6$&$~8.57$&$28.50$ & $0.20\pm0.05 $ & $12.86\pm0.74 $ & $0.18 $ & $0.08\pm0.02$ & $1.29\pm0.43$ & $~4.71\pm1.54$ & $3.66\pm0.75$ & -- \\             
      $~7$&$10.02$&$34.00$ & $0.34\pm0.05 $ & $19.97\pm0.76 $ & $0.32 $ & $0.16\pm0.03$ & $2.80\pm0.58$ & $~3.34\pm0.69$ & $2.99\pm0.38$ & -- \\             
      $~8$&$13.36$&$17.00$ & $0.15\pm0.03 $ & $12.30\pm0.68 $ & $0.14 $ & $0.08\pm0.02$ & $1.40\pm0.47$ & $~5.23\pm1.73$ & $3.91\pm0.81$ & [$37.44-49.32$]\\ 
      $~9$&$15.37$&$34.00$ & $0.23\pm0.05 $ & $14.25\pm0.67 $ & $0.21 $ & $0.05\pm0.01$ & $0.74\pm0.23$ & $~2.24\pm0.68$ & $2.31\pm0.44$ & [$5.52-7.28$] \\  
      $10$&$15.56$&$29.50$ & $0.27\pm0.06 $ & $17.59\pm1.02 $ & $0.26 $ & $0.11\pm0.03$ & $1.66\pm0.51$ & $~2.39\pm0.73$ & $2.42\pm0.46$ & [$2.53-3.34$] \\  
      $11$&$21.51$&$45.50$ & $0.72\pm0.11 $ & $26.66\pm1.03 $ & $0.68 $ & $0.32\pm0.06$ & $5.16\pm1.11$ & $~3.62\pm0.78$ & $3.14\pm0.42$ & -- \\             
      $12$&$22.48$&$32.00$ & $0.44\pm0.08 $ & $17.30\pm0.66 $ & $0.40 $ & $0.24\pm0.04$ & $4.73\pm0.86$ & $11.61\pm2.11$ & $6.43\pm0.73$ & ~[$8.32-10.96$]\\ 
      $13$&$33.95$&$26.50$ & $0.37\pm0.10 $ & $21.73\pm1.50 $ & $0.35 $ & $0.26\pm0.07$ & $5.06\pm1.64$ & $~4.92\pm1.60$ & $3.80\pm0.77$ & --\\          
        \hline
        \hline
        \end{tabular}
        \begin{tablenotes}\footnotesize
            \item[+] Core radius estimated from the initial maximum likelihood fitting.
            \item[$\ddagger$] Significance radius of the GCA and measured count-rate in this aperture.
            \item[$\ast$] $r_\textrm{500}$ and $CR_\textrm{500}$ are derived from the aperture count-rate combining the \cite{Reichert2011} scaling relations with a $\beta$-model of parameters $\beta=2/3$ and $r_\textrm{c}=r_{500}/7$.
            \item[$\dagger$] Interpolated values for the count-rate, flux, luminosity and mass within $r_\textrm{500}$, derived directly from the GCA measurement at $r_\textrm{500}$ and the \cite{Reichert2011} scaling relations (see Section \ref{sec:flux_estim} for details).
            \item[$\star$] Mass derived from the velocity dispersion for groups with more than $20$ galaxy members with spectroscopic information (see Section \ref{sec:m_sigmav} for details).
        \end{tablenotes}
    \end{threeparttable}
 \end{center}
\end{sidewaystable*}

\section{Notes on individual groups}
\label{sect:notes}

In this paper, we focus on 13 new X-ray selected galaxy groups that show very extended ($r_\textrm{sig}>10$~arcmin) emission in the RASS. Here we discuss each of them in detail.

Figure~\ref{fig:clustercandidates} shows each of these systems through the RASS photon images and optical images from ATLAS, SDSS or DSS, depending on the availability and in that order of importance. These images have $1\times1$~deg$^2$ size, except for the ATLAS images, which are $0.30\times0.30$~deg$^2$. The red cross shows the determined position of the detected group from the maximum likelihood fitting method, while the magenta plus signs display the positions of X-ray sources in the 1RXS and 2RXS catalogs. The green contours are obtained from the respective wavelet-filtered images, and the blue circle shows the core radius determined from the maximum likelihood fitting method (see Section~\ref{sec:method}). The cyan diamonds represent the positions of the individual galaxies with spectroscopic redshift information in NED and are used to determine the final redshift of the group (see Section~\ref{sec:redshift}).

As previously explained, the right panels of Figure~\ref{fig:clustercandidates} show the redshift determination and Table~\ref{tab:candi2} shows a summary of the parameters estimated for the $13$ systems (see Section~\ref{sec:newclu}). We refer the reader to the right panels of Figure~\ref{fig:clustercandidates} and Table~\ref{tab:candi} while reading the notes on the individual groups.

\subsection{Group no.~$1$}
\label{sec:clu1}

NED provides redshift information for only $6$ galaxies in this group at $z\sim0.07$. The SDSS image (see Figure~\ref{fig:1a}) shows two clear and prominent brightest cluster galaxies (BCGs) at the center of our X-ray detection, for which there is no spectroscopic information in NED. From SDSS~DR14, we found photometric redshifts of $0.059$ and $0.068$ for these central galaxies, which are consistent with our redshift estimation for this group ($z=0.069$). This system is the most compact group in our pilot sample ($r_{500}\sim8$~arcmin).

\subsection{Group no.~$2$}
\label{sec:clu2}

According to NED, the group no.~$2$ has been previously identified as a galaxy group. Within $10$~arcmin of the group position, two of these previous detections are: USGC~S110 located $2.17$~arcmin from the group center at $z=0.012$ (\citealt{Ramella2002}), and source~$425$ of \citealt{Diaz-Gimenez2015} positioned $\sim 8.37$~arcmin away at $z=0.013$. These redshifts are consistent with our estimation for this group ($z=0.013$ from $10$ galaxies). The ATLAS image shows two main dominating galaxies at center (see Figure~\ref{fig:1b}). This object is the most extended ($r_{500}\sim30$~arcmin) and nearest group in our pilot sample. It has the largest flux of all groups in the sample ($F_{500}^\textrm{interp}\sim7.74\times10^{-12}~$erg~s$^{-1}$~cm$^{-2}$), which is $2.6$~times larger than the flux limit of REFLEX catalog ($3\times10^{-12}$~erg~s$^{-1}$~cm$^{-2}$ in the $0.1-2.4$~keV energy band, \citealt{Boehringer2001}).

\subsection{Group no.~$3$}

This group has been previously identified in several publications, for example, as SDSS-C4~1156 at $z=0.039$ by \citet{Linden2007}, and as PM2GC~GR1204 at $z=0.040$ by \citet{Calvi2011}. These works provided several spectroscopic redshifts (for $42$~galaxies), which allow us to obtain a redshift estimation of $z=0.039$. Moreover, this group has several elliptical galaxies easily spotted by eye (see Figure~\ref{fig:1c}).

\subsection{Group no.~$4$}

Group no.~$4$ presents a similar situation with group no.~$1$ (see Section~\ref{sec:clu1}). NED provides redshift information for only $7$ galaxies, obtaining an estimated redshift $z=0.072$ for this system. From the 2MASS Photometric Redshift catalog (2MPZ, \citealt{Bilicki2014}), we found photometric redshifts of $22$ galaxies. These redshift values are consistent with the final redshift estimation of this galaxy group. The spatial distribution of these galaxies coincides with the X-ray contours of this system. This group has the highest redshift value in our pilot sample. Although this object is located nearby a bright point-like source and surrounded by another two weak RXS point-like sources, our detection algorithm is capable to disentangle all this information and it still finds this extended source. The DSS image shows a bright galaxy near the X-ray peak emission of the group (see Figure~\ref{fig:1d}). This system is the most massive ($M_{500}^\textrm{interp}\sim8.36\times10^{13}~$M$_\odot$) and most luminous ($L_{500}^\textrm{interp}\sim18.25\times10^{42}$~erg~s$^{-1}$) group in our pilot sample.

\subsection{Group no.~$5$}
\label{sec:clu5}

Similar to group no.~$2$ (see Section~\ref{sec:clu2}), this system has been identified in different works: \citet{Colless2001} (identified as CID~15 at $z=0.036$ and located $5.04$~arcmin away from the group center), and \citet{Ramella2002} (labeled as USGC~U118 at $z=0.035$ and located $8.64$~arcmin away). Both optical detections have similar redshifts to the one determined from $17$ galaxy members ($z=0.036$). The DSS image (see Figure~\ref{fig:1e}) clearly shows elliptical galaxies in this field.
 
\subsection{Group no.~$6$}

This group has also been identified as a galaxy triplet system, APMUKS(BJ)~B013430.06-14130 \citep{Maddox1990}, located $2.89$~arcmin away from the group center, as a galaxy group USGC~S061 at $z=0.039$, $6.70$~arcmin away \citep{Ramella2002}, and as a galaxy group named [DZ2015]~396 at $z=0.039$ located $7.26$~arcmin away \citep{Diaz-Gimenez2015}. Although NED only provides redshifts for $14$ galaxies, the estimated redshift ($z=0.040$) of the group is consistent with the ones of the identified close groups. The ATLAS image (Figure~\ref{fig:1f}) clearly shows a dominant galaxy in the system with $z=0.040$ \citep{Huchra1993}.

\subsection{Group no.~$7$}

In a similar situation as group no.~$5$ (see Section~\ref{sec:clu5}), NED shows an identified optical cluster counterpart $1.92$~arcmin away from this group. This matching is called WBL~704 at $z=0.024$ (\citealt{White1999}). This redshift is consistent with our estimation for this group ($z=0.023$ from $8$ galaxies). The SDSS image shows a dominant BCG (see Figure~\ref{fig:1g}).

\subsection{Group no.~$8$}
\label{sec:clu8}

For this system, there are $123$ galaxies with redshift measured from NED, giving $z=0.040$ as the final result. It has been previously reported by \citet{Smith2012} as MSPM~00022 located $1.94$~arcmin away from the group center at $z=0.039$, which is consistent with our redshift determination. The SDSS image shows many elliptical galaxies gathering around a prominent BCG (Figure~\ref{fig:1h}). This group is located $47.96$~arcmin away from cluster Abell~2152 at $z=0.041$ \citep{Struble1999} with $r_{500}=12.99$~arcmin \citep{Piffaretti2011}. Given that our group presents a similar redshift, we might consider it as a potential infalling group into Abell~2152. This might be reflected by the large discrepancy between its $M_{500}^\textrm{interp}$ ($\sim 3.91\times10^{13}~$M$_\odot$) and $M_{500}^{\sigma_\textrm{v}}$ ($\sim 37.44\times10^{13}~$M$_\odot$) values. However, further investigation is needed. 

\subsection{Group no.~$9$}

 Similar to group no.~$5$ (see Section~\ref{sec:clu5}), this system has an optical counterpart, MSPM~00069 (\citealt{Smith2012}) located $5.50$~arcmin away at $z=0.035$, which is consistent with our redshift estimation for this group ($z=0.036$ from $52$ galaxies). ROSAT catalogs and image (Figure~\ref{fig:1i}) show that this group is surrounded by point-like sources, but our detection algorithm is able to disentangle the X-ray emission and detect them individually. This group is the least massive ($M_{500}^\textrm{interp}\sim2.31\times10^{13}~$M$_\odot$), the least luminous ($L_{500}^\textrm{interp}\sim2.24\times10^{42}$~erg~s$^{-1}$) and it has the smallest flux ($F_{500}^\textrm{interp}\sim0.74\times10^{-12}~$erg~s$^{-1}$~cm$^{-2}$) of all groups in the sample.

\subsection{Group no.~$10$}

This system has been reported in previous works as HCG~067 (\citealt{Hickson1982}) and as LDCE~1002 (\citealt{Crook2008}), at redshifts $0.025$ and $0.024$, respectively. These redshifts are consistent with the one we have estimated for this group ($z=0.025$ from $22$ galaxies). Although this group is surrounded by three bright RXS sources (Figure~\ref{fig:1j}), our detection algorithm is able to disentangle the individual emission. The ATLAS image clearly shows several elliptical galaxies in this system.

\subsection{Group no.~$11$}

Within $10$~arcmin of the position of this group, we found two Abell clusters: Abell~S0924 at $z=0.016$ and Abell~3742 at $z=0.016$ \citep{Abell1989}. These redshifts are fairly consistent with our redshift estimation of $z=0.017$ for this group, which was obtained from $10$~galaxies. Given that this group displays two prominent BCGs (Figure~\ref{fig:1k}), it has been identified in other works as a galaxy group at a redshift similar to our estimation (e.g., \citealt{Chow-Martinez2014}). 

\subsection{Group no.~$12$}

From $38$ galaxy redshifts, we determined $z=0.033$ for this system. The DSS image (see Figure~\ref{fig:1l}) shows several elliptical galaxies in the area of this group. It coincides with several kown optical counterparts. The most representative are groups~$266$ of \citealt{Diaz-Gimenez2015} and $77$ of \citet{Ragone2006}, both with a redshift of $z=0.034$ and located respectively $5.44$ and ${\sim}8$~arcmin from our detection. In a similar situation as for group no.~$8$, it is located only $27.00$~arcmin away from galaxy cluster Abell~S$0736$, also at $z=0.033$ \citep{Chow2014}, and $58.95$~arcmin from galaxy cluster Abell~$1736$ at $z=0.046$ \citep{Planck2016a}. This group is actually a member of the Front-Eastern-Wall of the Shapley supercluster. According to the groups identified in the area by \citet{Ragone2006}, it most likely resides in a filamentary structure starting from Abell~S$0736$ ($z=0.033$), going through our source ($z=0.033$), then Abell~$1736$ NED01 ($z=0.035$), Abell~$1736$ NED02 ($z=0.041$) and ending at Abell~$1736$ ($z=0.0453$), all nicely aligned also in the plane of the sky. This is further supported by their location in the outskirts of the Shapley supercluster (they have some of the lowest redshift among the identified supercluster members). 

\subsection{Group no.~$13$}

Based on $14$ redshifts found in NED, we determine $z=0.021$ for this group. It has three optical counterparts at similar redshift within $5$~arcmin: galaxy pair NGC~1044 (\citealt{Falco1999}), galaxy triple HDCE~0161 (\citealt{Crook2008}), and galaxy group USGC~U142 (\citealt{Ramella2002}). The DSS image (see Figure~\ref{fig:1m}) shows a galaxy pair dominating the system.

\section{Discussion}

The $13$ groups in our pilot sample have extent values (i.e., core radius) larger than $5$~arcmin. Out of these $13$ groups, $11$ have a redshift lower than $0.05$. For all of them, significant emission can be traced out to $r_\textrm{sig}>10$~arcmin (see Table~\ref{tab:candi}). These facts show that indeed some local, very extended galaxy groups have been missed in previous works and that our methodology used to look for them works fine.

The flux measurements of our groups show that these $13$ systems are above the flux limit of the 1RXS and 2RXS catalogs ($\sim10^{-13}$~erg~s$^{-1}$~cm$^{-2}$ for 2RXS catalog, \citealt{Boller2016}). Since most RASS galaxy cluster catalogs rely on 1RXS/2RXS as initial source sample, it is not surprising the galaxy groups are also not included in any published RASS cluster catalog, despite their fluxes often being larger than the flux limits of those. 

Our growth curve analysis method shows that our groups have a relatively flat ($\beta<2/3$) surface brightness distribution (see the third panel of each row in Figure~\ref{fig:clustercandidates}), making this pilot sample a perfect representation of sources that do not have the expected/usual characteristics and therefore are missed in catalogs. This is supported by the locations of the $13$ groups in the extension likelihood - extent plane (see right panel of Figure~\ref{fig:extextlikeplane}), which are consistent with detected clusters of low $\beta$-values in our simulations (low values of extension likelihood and large values of extent, see Figure~\ref{fig:diffbeta_Sel}).

All these features make our $13$ groups perfect candidates to be missed by the sliding-cell algorithm used to construct the RASS catalogs. This source finder performs best on detecting point-like sources \citep{Rosati2002}. 

The fact that most of our secure group candidates lie below $z=0.05$ raises the question whether there is a significant fraction of such systems that have been missed. Around $14\%$ of the MCXC clusters, $251$ have redshift no larger than $z=0.05$. A similar number, $219$ objects, constitutes our sample of group candidates with no 1RXS/2RXS detections. If further analysis shows that at least $50\%$ of these systems are indeed galaxy groups we will increase the number of known X-ray galaxy groups at $z<0.05$ by at least $40\%$. \citealt{Schellenberger2017} have shown the impact on the values of $\Omega_\textrm{m}$ and $\sigma_8$ by the incompleteness of the mass function in the galaxy group regime. Basically, if there is an uncorrected incompleteness of $25\%-50\%$ at low masses, the best-fit $\Omega_\textrm{m}$ values are biased low.

\section{Conclusions}
\label{sec:con}

In this work, we have reprocessed the ROSAT All-Sky Survey (RASS) data with the aim to look for very extended, low surface brightness galaxy groups and clusters that might have been missed in previous works. Our approach consisted of a wavelet-based source detection algorithm and a maximum-likelihood fitting method to characterize the source properties. We tested our methodology on extensive Monte-Carlo simulations and determined the selection criteria for significant and extended sources.

We found $1\,308$ cluster candidates in the RASS data. $\sim44\%$ of those systems have a cluster counterpart in either MCXC or PSZ2 catalogs. This makes our methodology a reliable procedure. From the $\sim56\%$ remaining, $70\%$ of them have a counterpart in either the 1RXS or the 2RXS catalogs. In this paper, we focus on and perform an extensive visual inspection and literature search of the other $30\%$ of the systems. Out of these objects, we describe in detail the analysis and characterization of a pilot sample of $13$ candidates.

We determined a spectroscopic redshift for each cluster candidate finding that most of them are local systems ($z\lesssim0.05$), and are very extended (significant X-ray emission traced out to more than $10$~arcmin radius). Moreover, all these objects have fluxes above the limit of the 1RXS and 2RXS catalogs. We argue that they might have been missed because of the used source finder in previous works, sliding-cell algorithm, which is not optimal for finding such low flux, very extended objects. Our pilot sample also reveals that the surface brightness distribution of the systems is flatter than the usual profile, i.e.\ $\beta<2/3$, and that the missed systems are located in the group regime given the determined masses from galaxy velocity dispersions or a $L_\textrm{X}$-$M$ scaling relation ($M_{500}\lesssim10^{14}~$M$_{\odot}$).

In a subsequent paper, we will present and discuss in detail the remaining cluster candidates that we have found.

\begin{acknowledgements}
The authors wish to thank Angus Wright, Jens Erler, Chaoli Zhang, Zhonglue Wen and Jinlin Han for useful help and discussions during the development of this paper. This work was supported by the German Research Association (DFG) through the Transregional Collaborative Research Centre TRR33 The Dark Universe (project B18) and the German Aerospace Agency (DLR) with funds from the Ministry of Economy and Technology (BMWi) through grant 50 OR 1514. The authors thank the support from the CAS-DAAD Joint Fellowship Programme for Doctoral Students of Chinese Academy of Sciences (ST 34). WX acknowledges the support of the Chinese Academy of Sciences through grant No. XDB23040100 from the Strategic Priority Research Program and that of the National Natural Science Foundation of China with grant No. 11333005. This research has made use of the NASA/IPAC Extragalactic Database (NED), which is operated by the Jet Propulsion Laboratory, California Institute of Technology, under contract with the National Aeronautics and Space Administration.
\end{acknowledgements}

\bibliographystyle{aa} 
\bibliography{main.bbl} %ref.bib} 

\begin{sidewaysfigure*}[ht]
\centering
\begin{subfigure}{\textwidth}
\centering
\includegraphics[width=0.24\hsize]{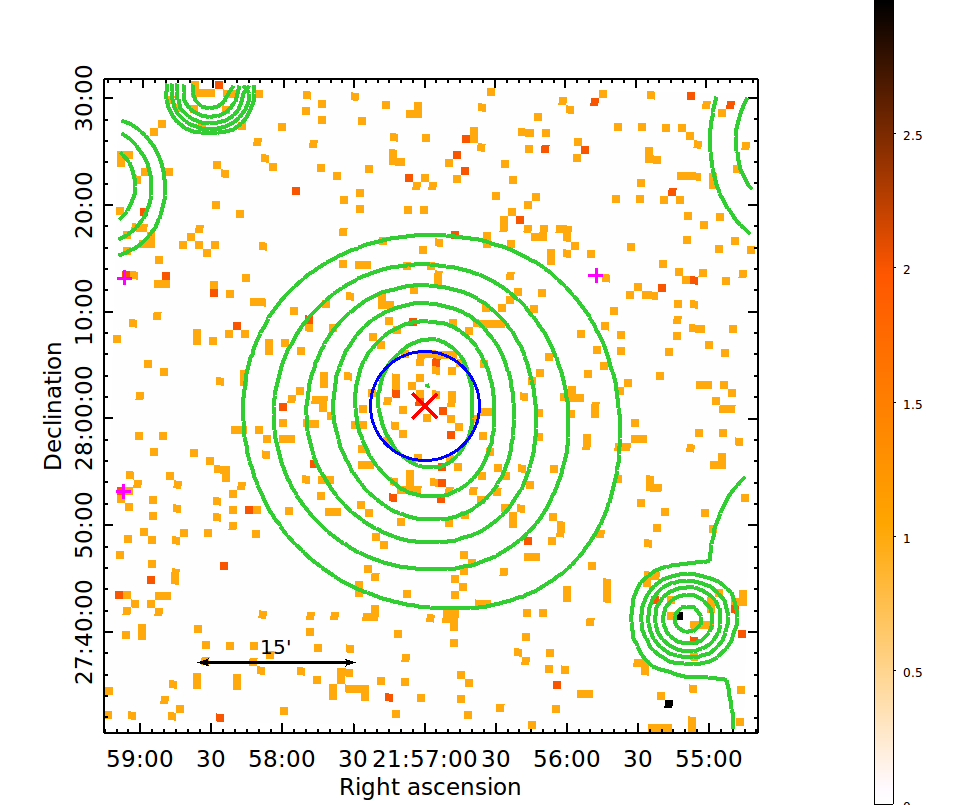}
\includegraphics[width=0.2\hsize]{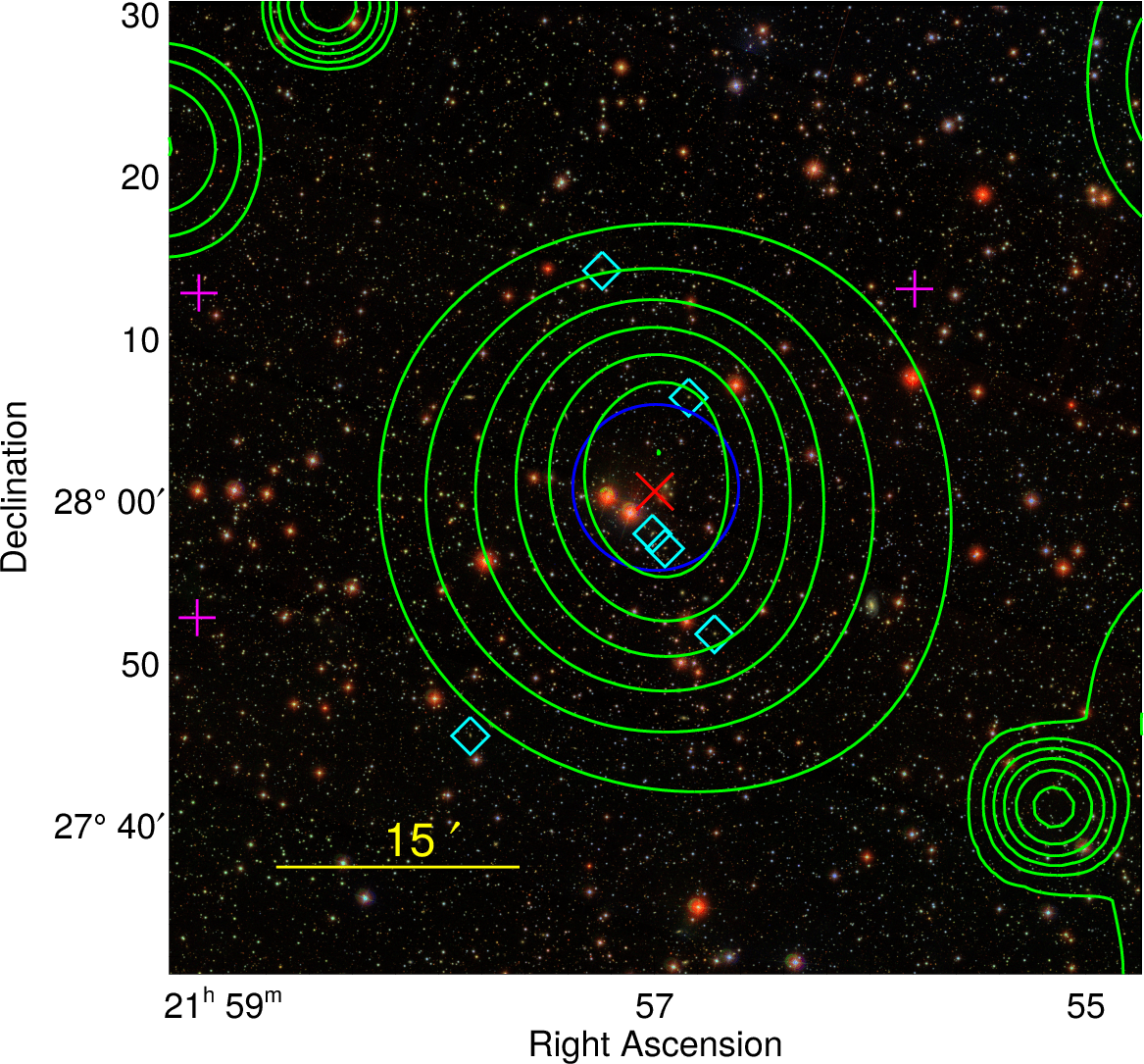}
\includegraphics[width=0.24\hsize]{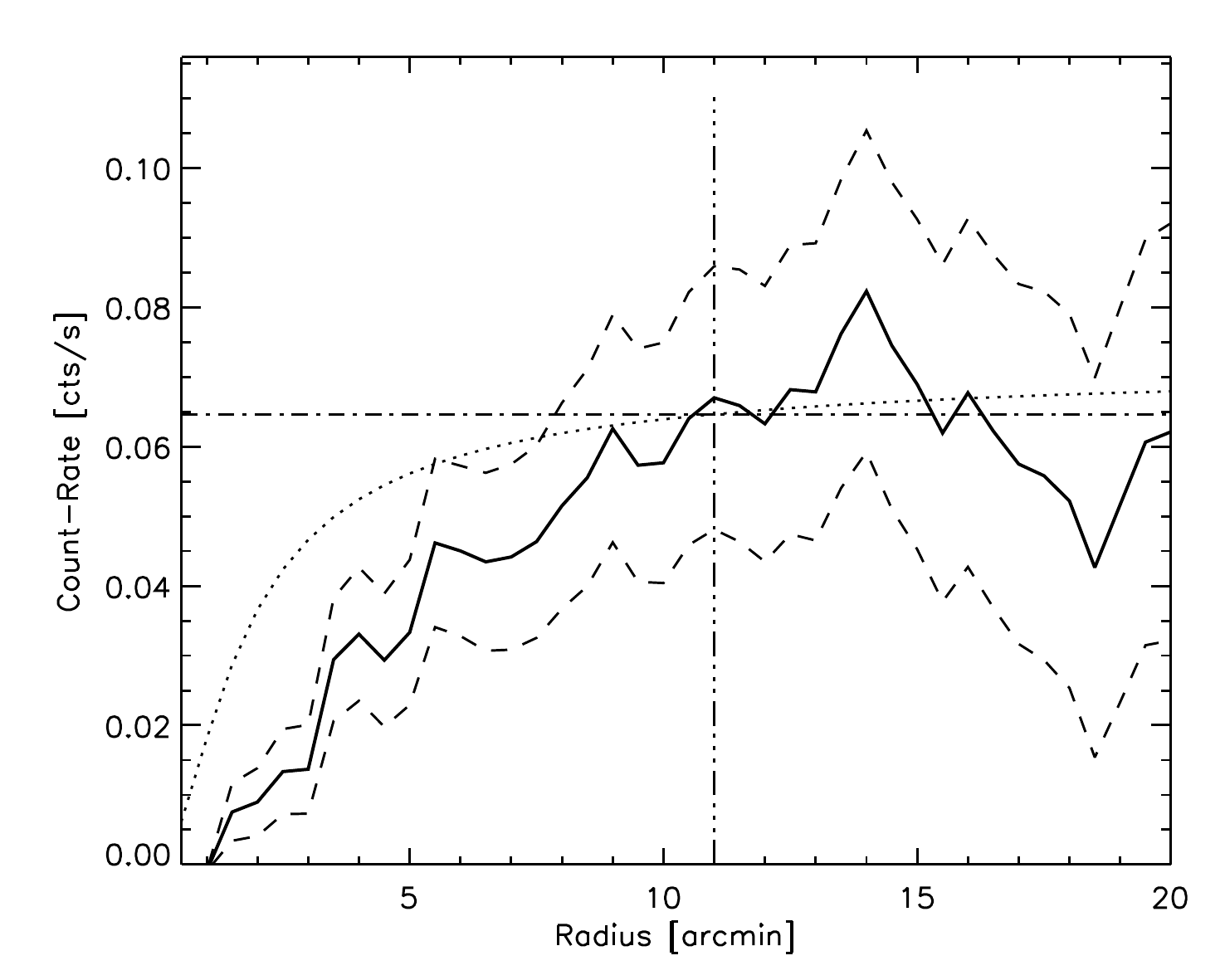}
\includegraphics[width=0.27\hsize]{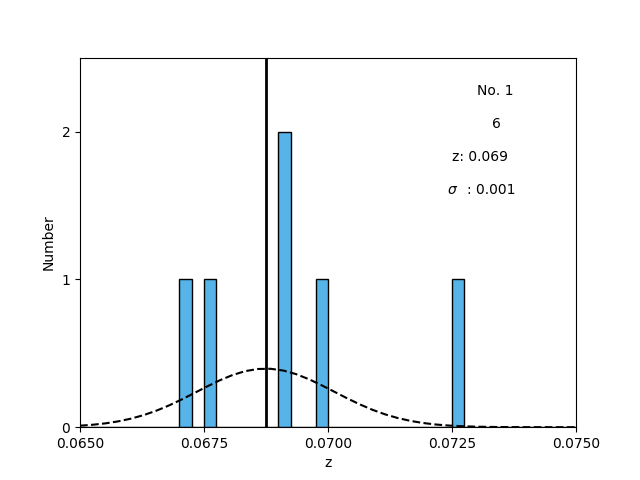}
\caption{\footnotesize{Group no.~$1$}} \label{fig:1a}
\end{subfigure}
\hspace*{\fill}
\begin{subfigure}{\textwidth}
\centering
\includegraphics[width=0.24\hsize]{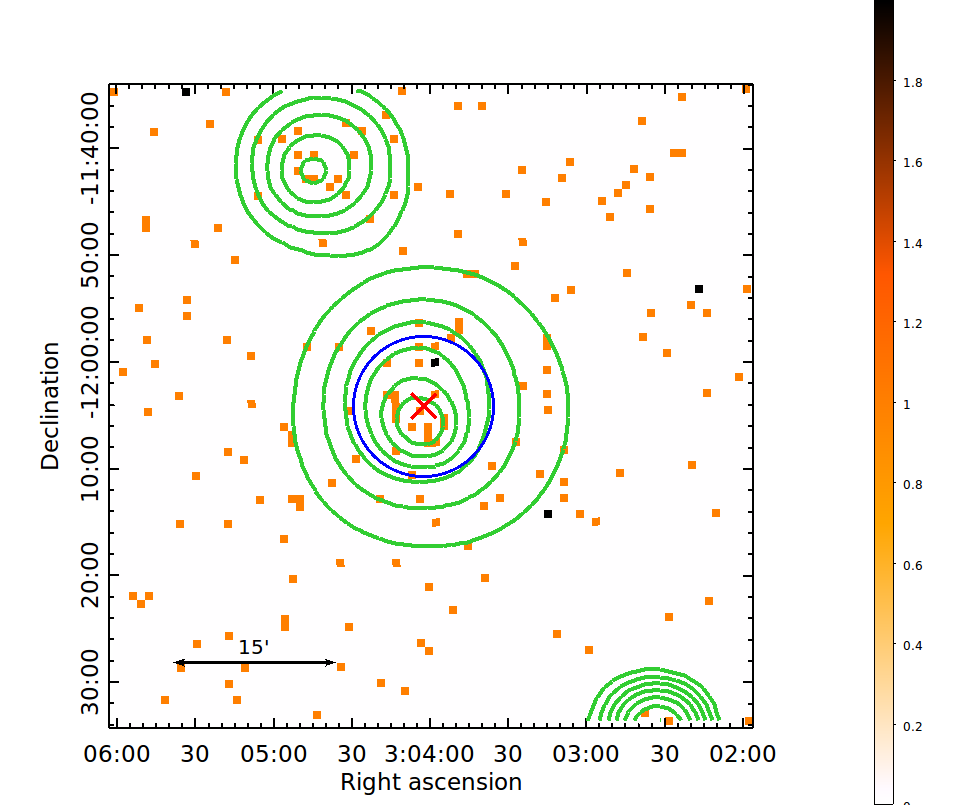}
\includegraphics[width=0.195\hsize]{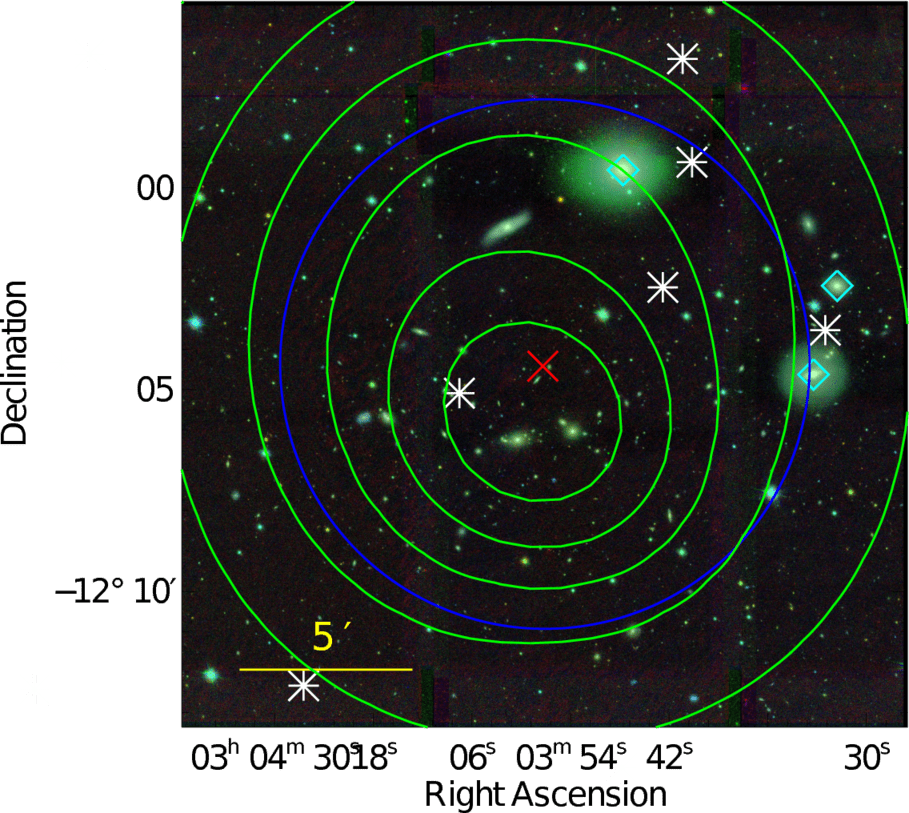}
\includegraphics[width=0.24\hsize]{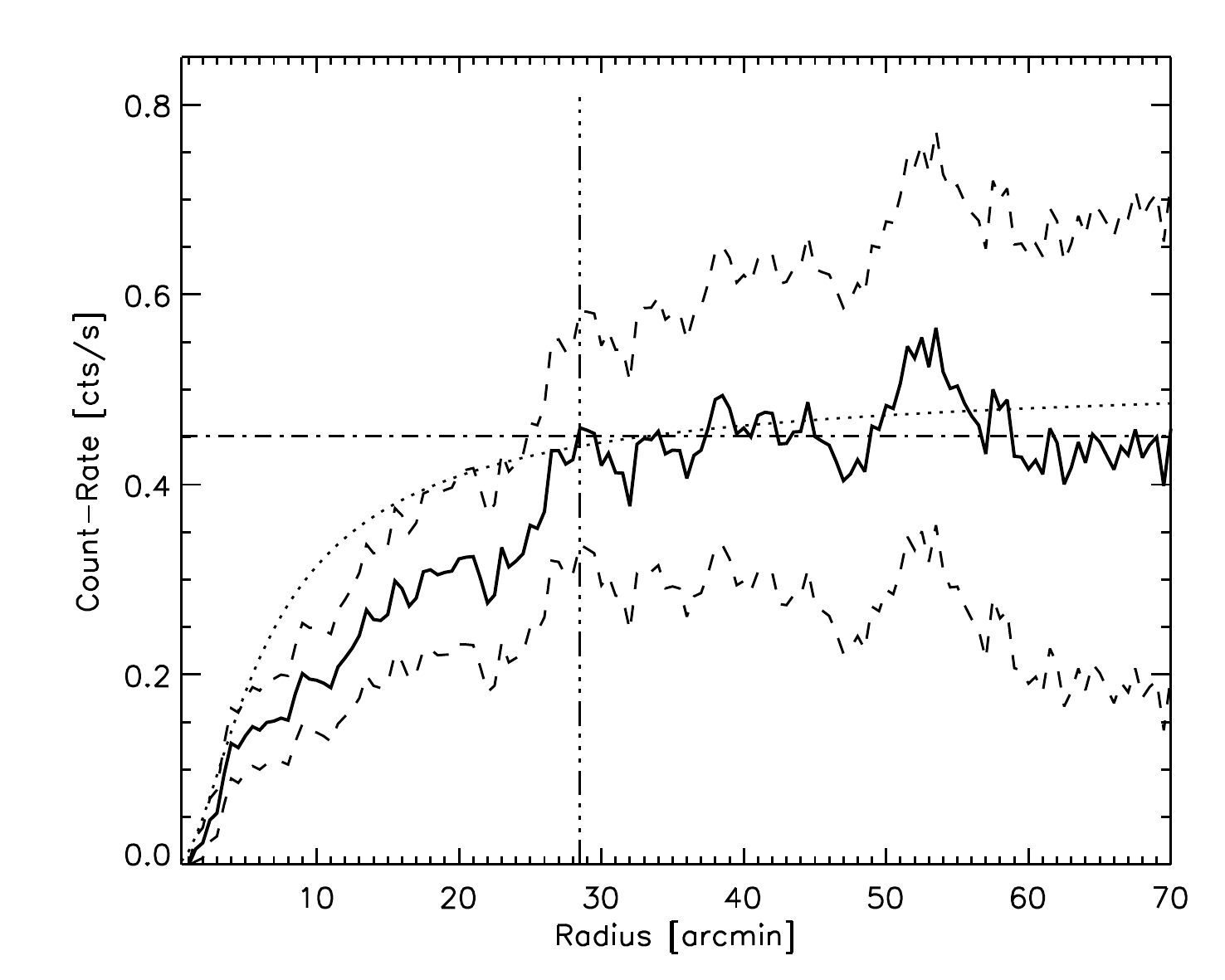}
\includegraphics[width=0.27\hsize]{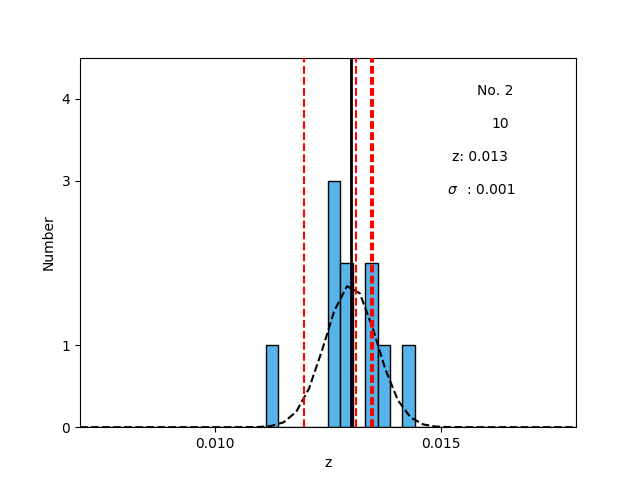}
\caption{\footnotesize{Group no.~$2$}} \label{fig:1b}
\end{subfigure}
\hspace*{\fill}
\begin{subfigure}{\textwidth}
\centering
\includegraphics[width=0.24\hsize]{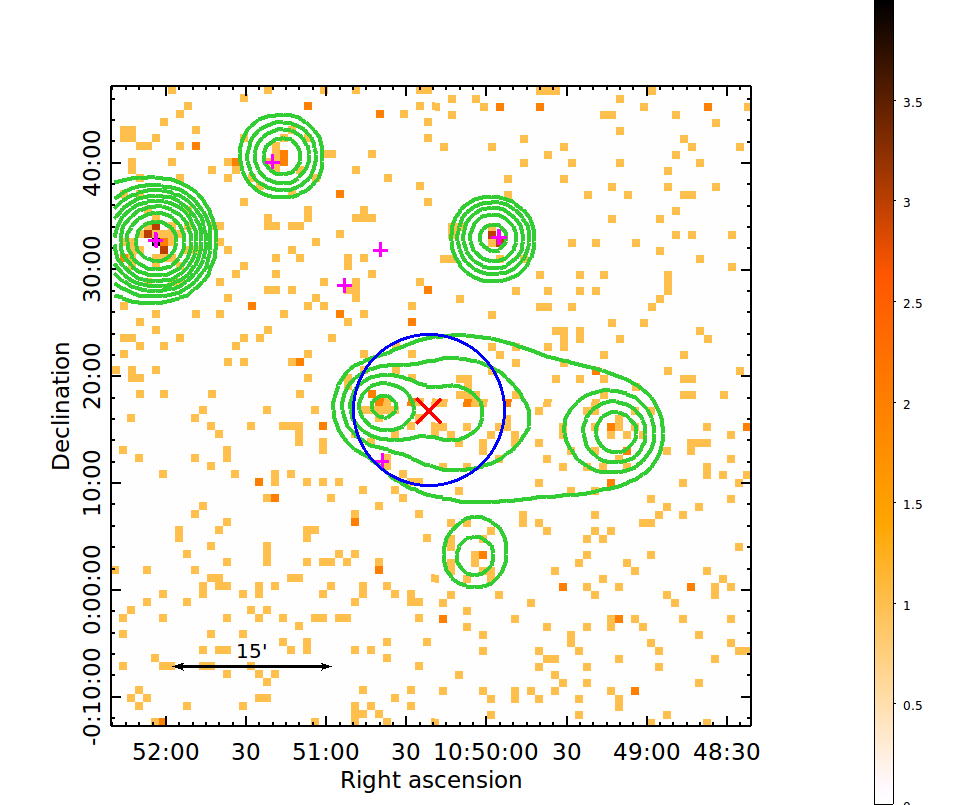}
\includegraphics[width=0.2\hsize]{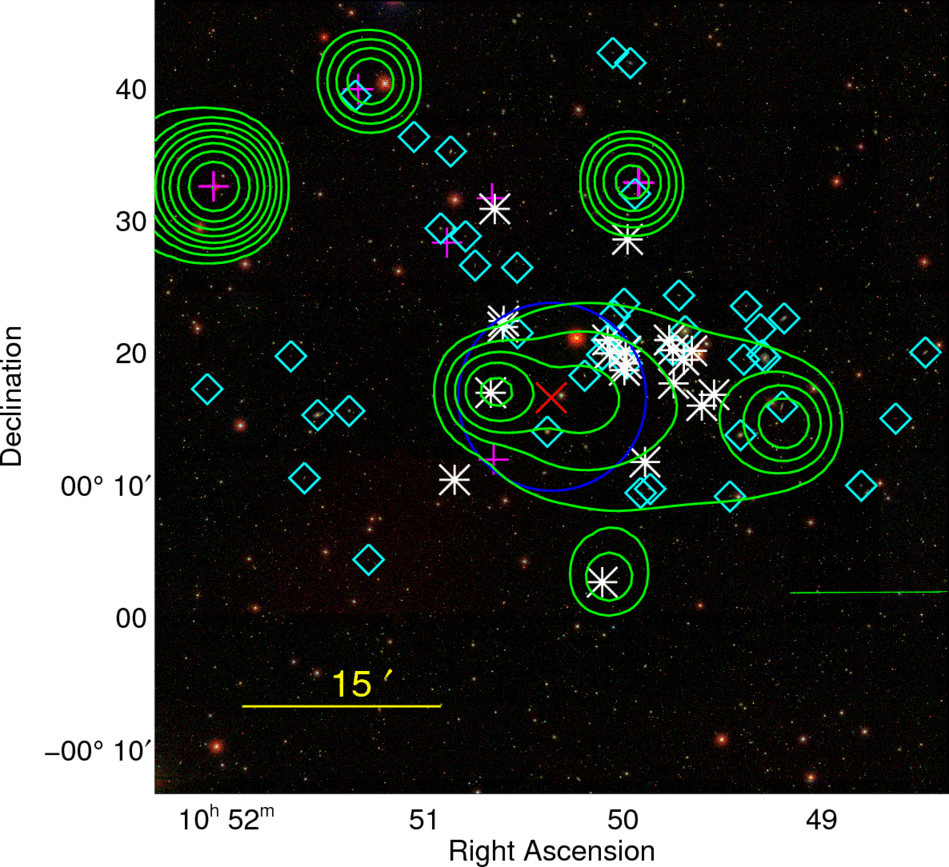}
\includegraphics[width=0.24\hsize]{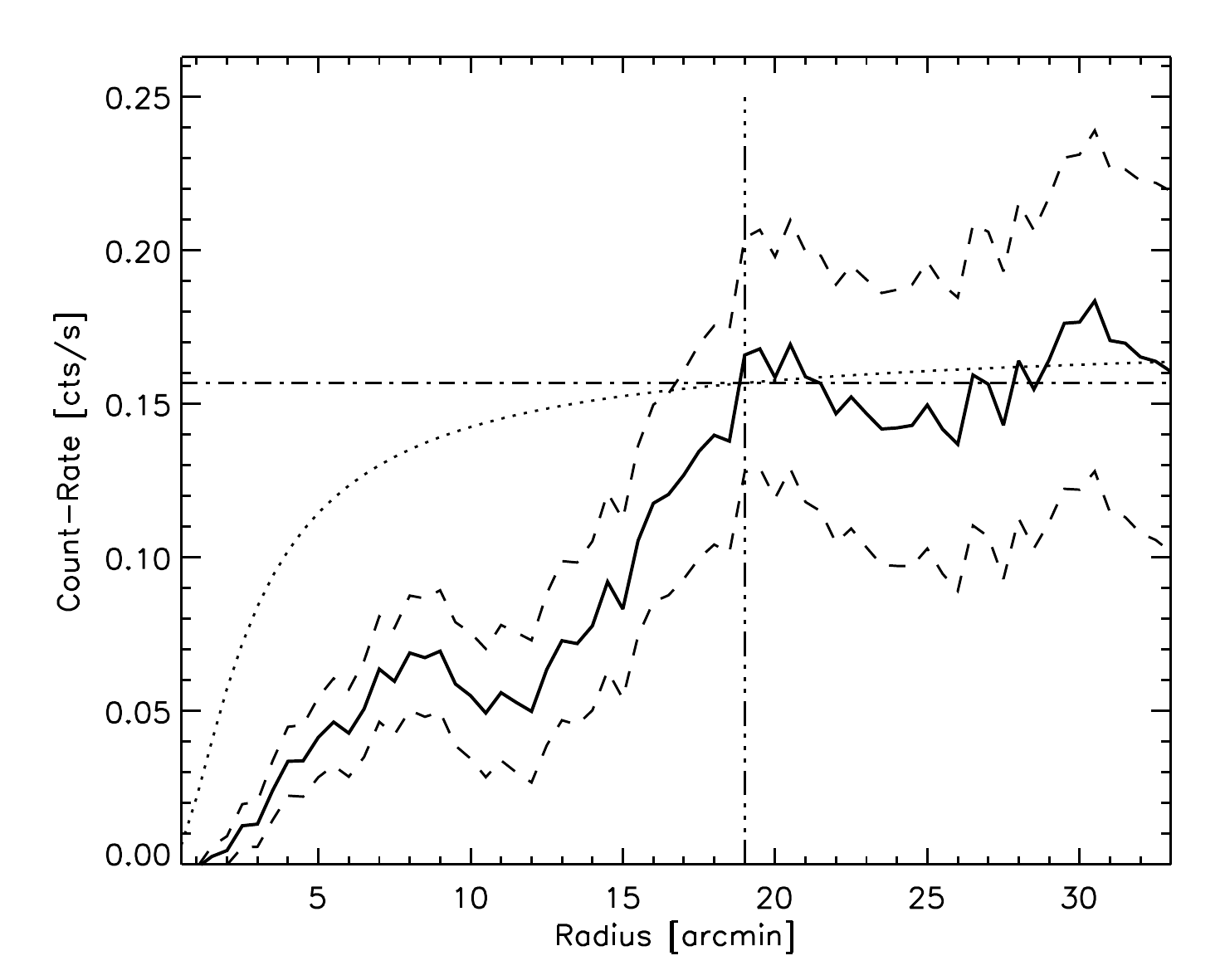}
\includegraphics[width=0.27\hsize]{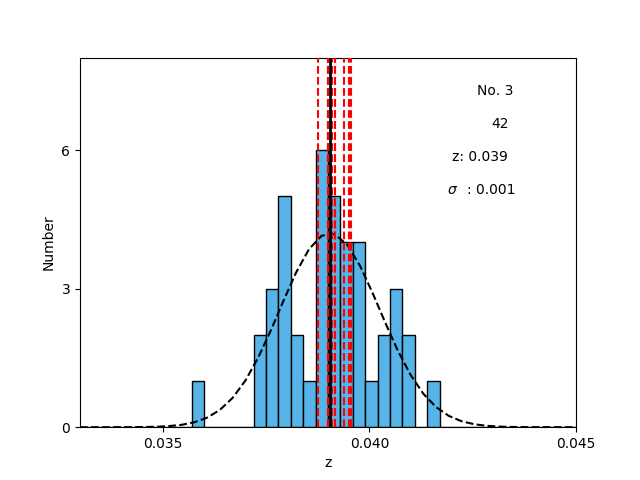}
\caption{\footnotesize{Group no.~$3$}} \label{fig:1c}
\end{subfigure}
\end{sidewaysfigure*}

\newpage

\begin{sidewaysfigure*}
\centering
\ContinuedFloat
\hspace*{\fill}
\begin{subfigure}{\textwidth}
\centering
\includegraphics[width=0.24\columnwidth]{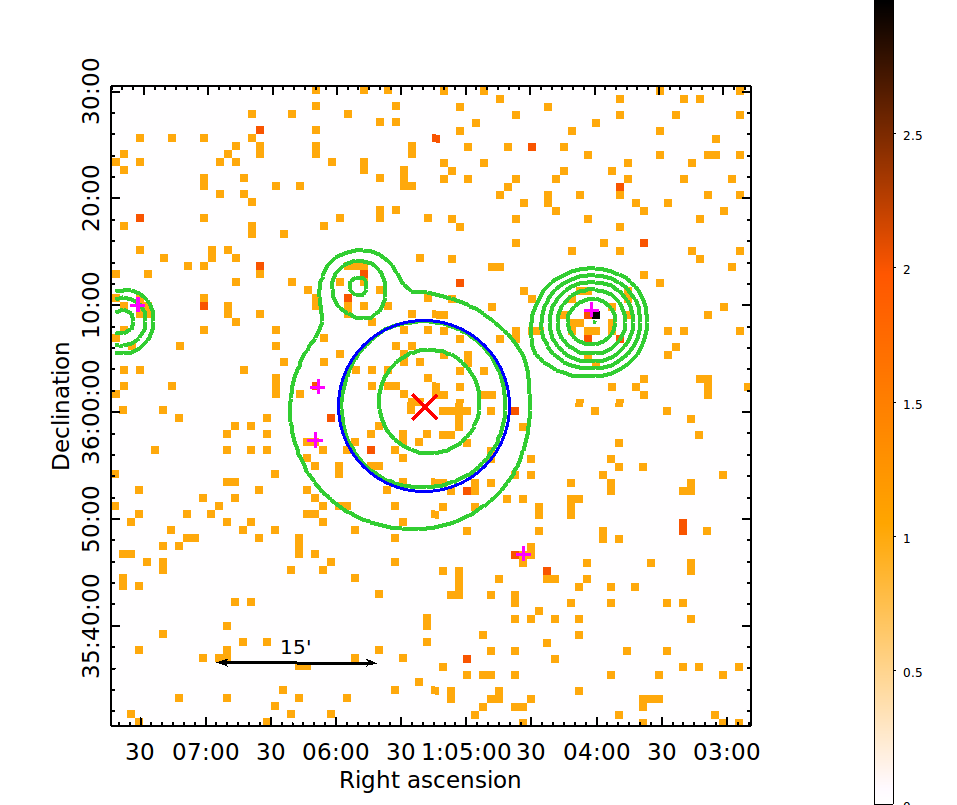}
\includegraphics[width=0.2\columnwidth]{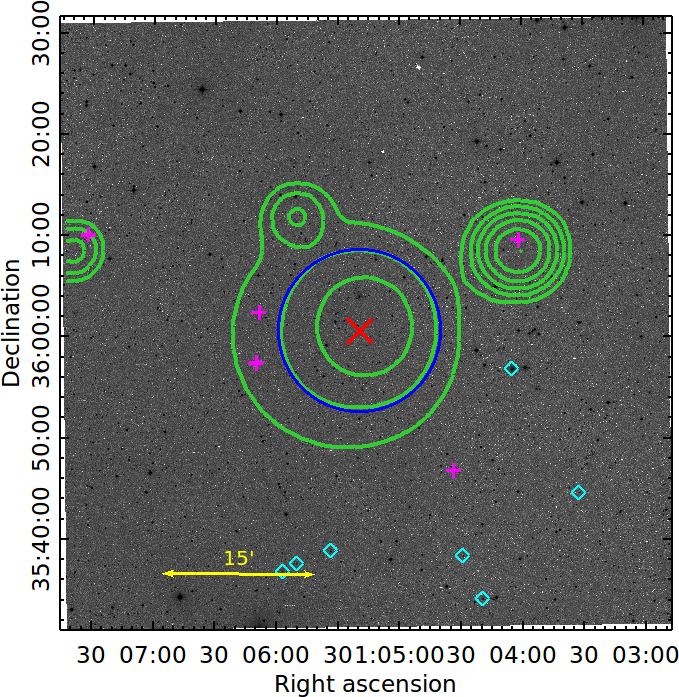}
\includegraphics[width=0.24\linewidth]{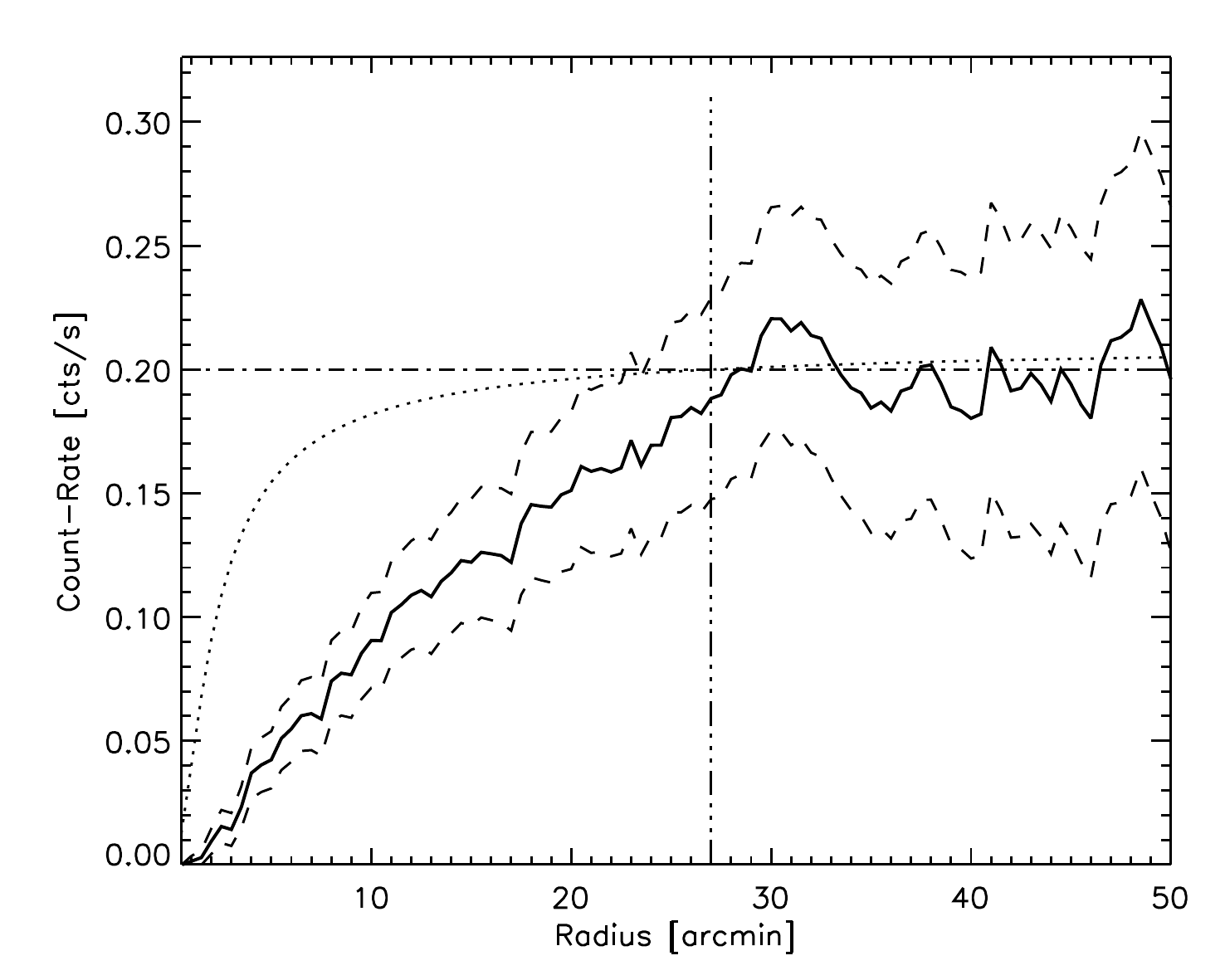}
\includegraphics[width=0.27\hsize]{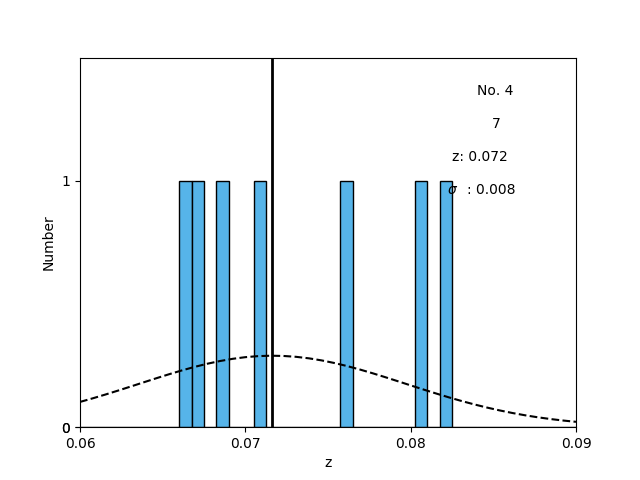}
\caption{\footnotesize{Group no.~$4$}} \label{fig:1d}
\end{subfigure}
\hspace*{\fill}
\begin{subfigure}{\textwidth}
\centering
\includegraphics[width=0.24\columnwidth]{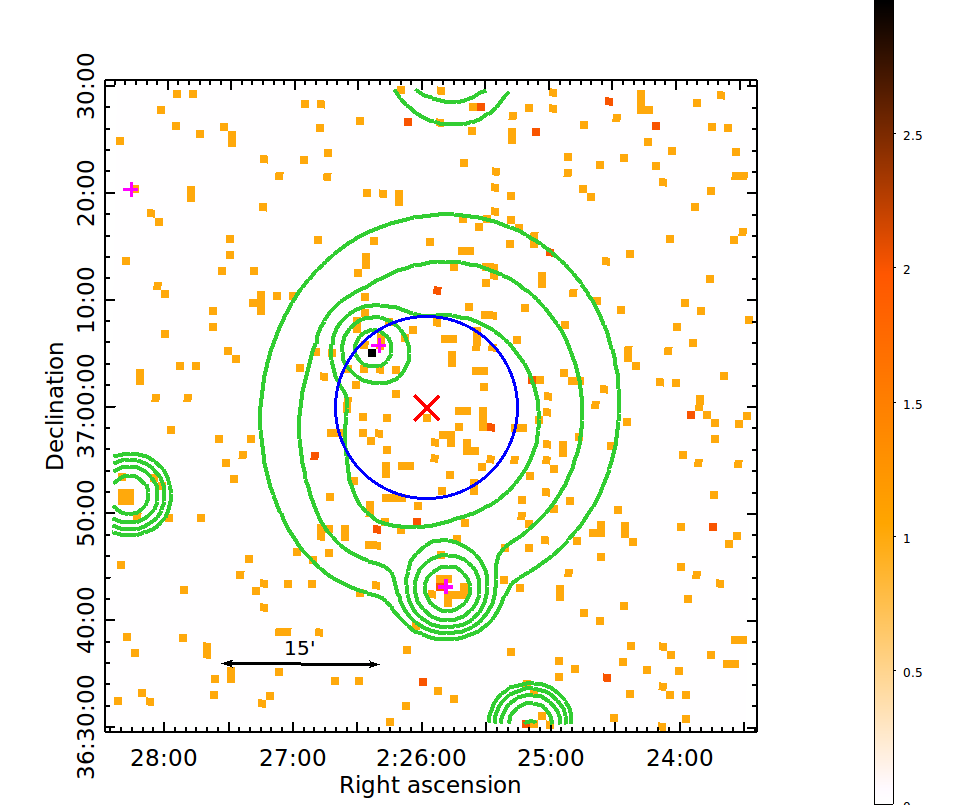}
\includegraphics[width=0.2\columnwidth]{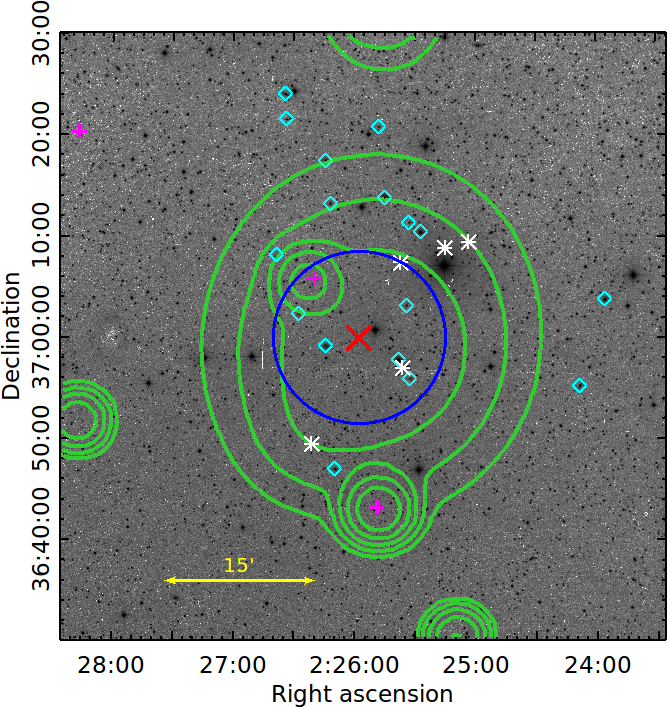}
\includegraphics[width=0.24\linewidth]{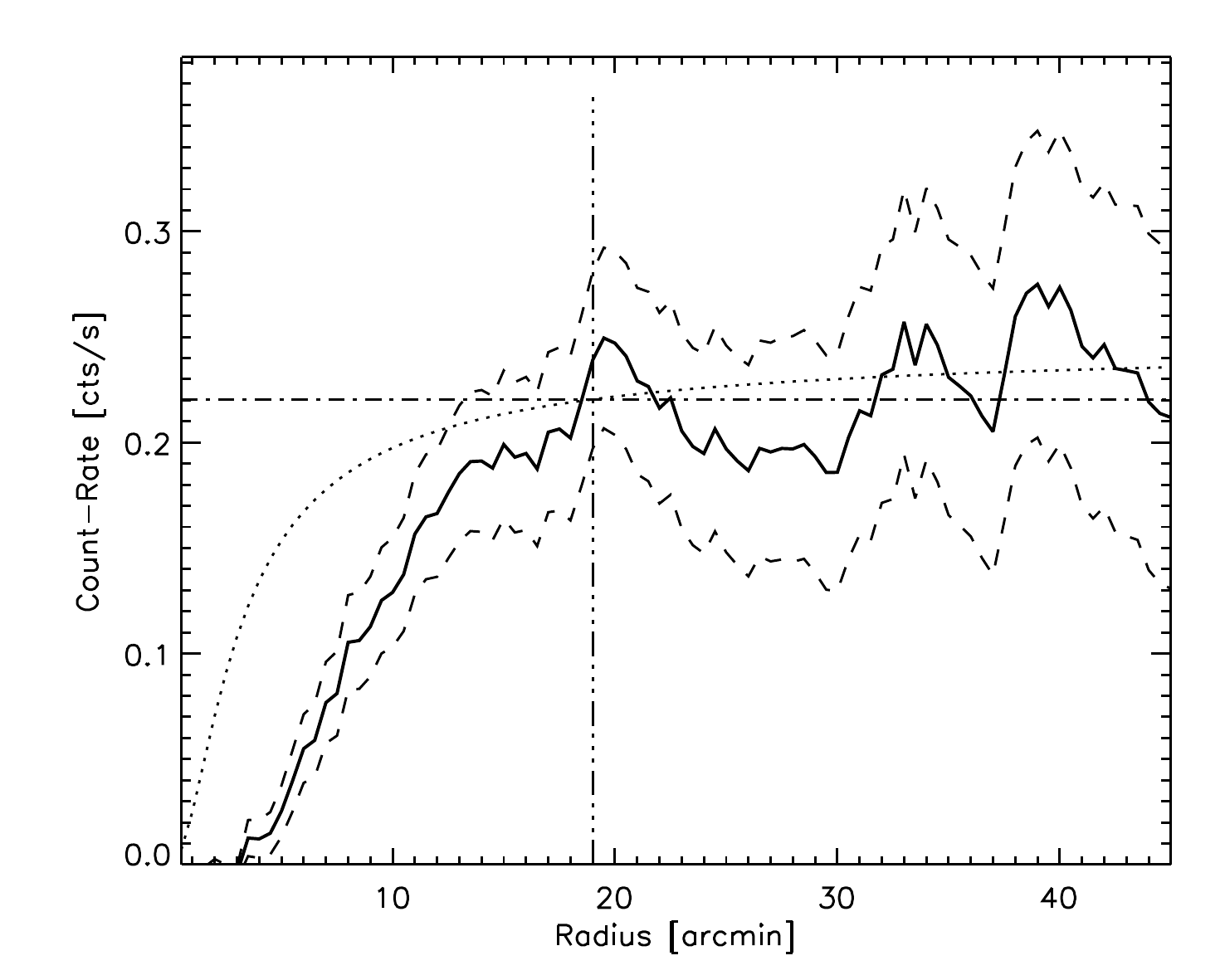}
\includegraphics[width=0.27\hsize]{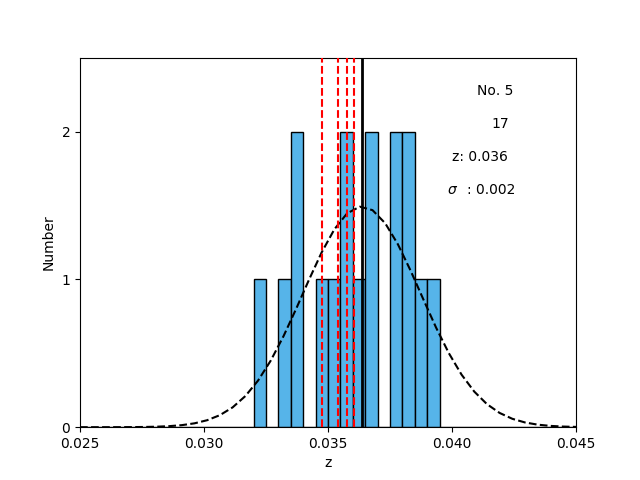}
\caption{\footnotesize{Group no.~$5$}} \label{fig:1e}
\end{subfigure}
\hspace*{\fill}
\begin{subfigure}{\textwidth}
\centering
\includegraphics[width=0.24\columnwidth]{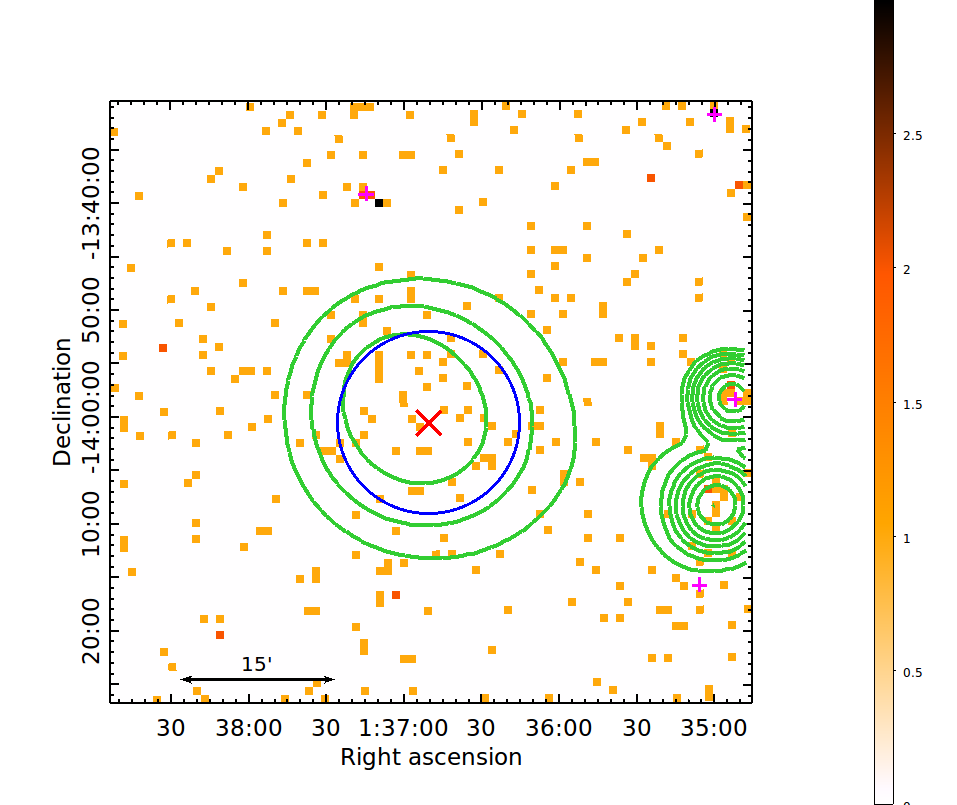}
\includegraphics[width=0.195\columnwidth]{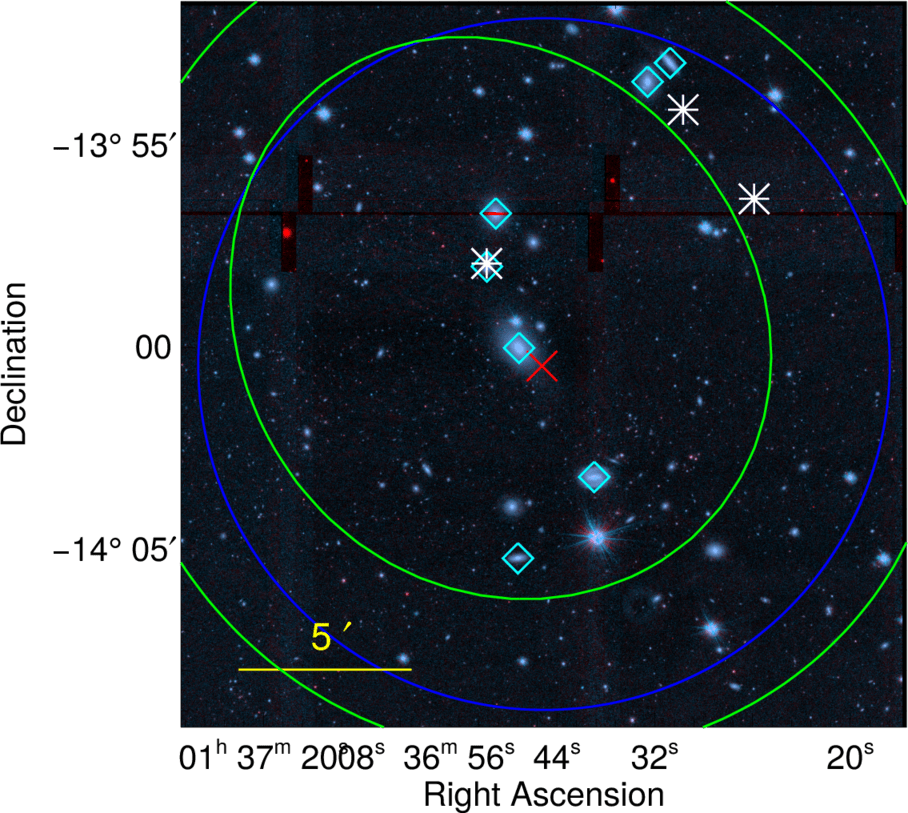}
\includegraphics[width=0.24\linewidth]{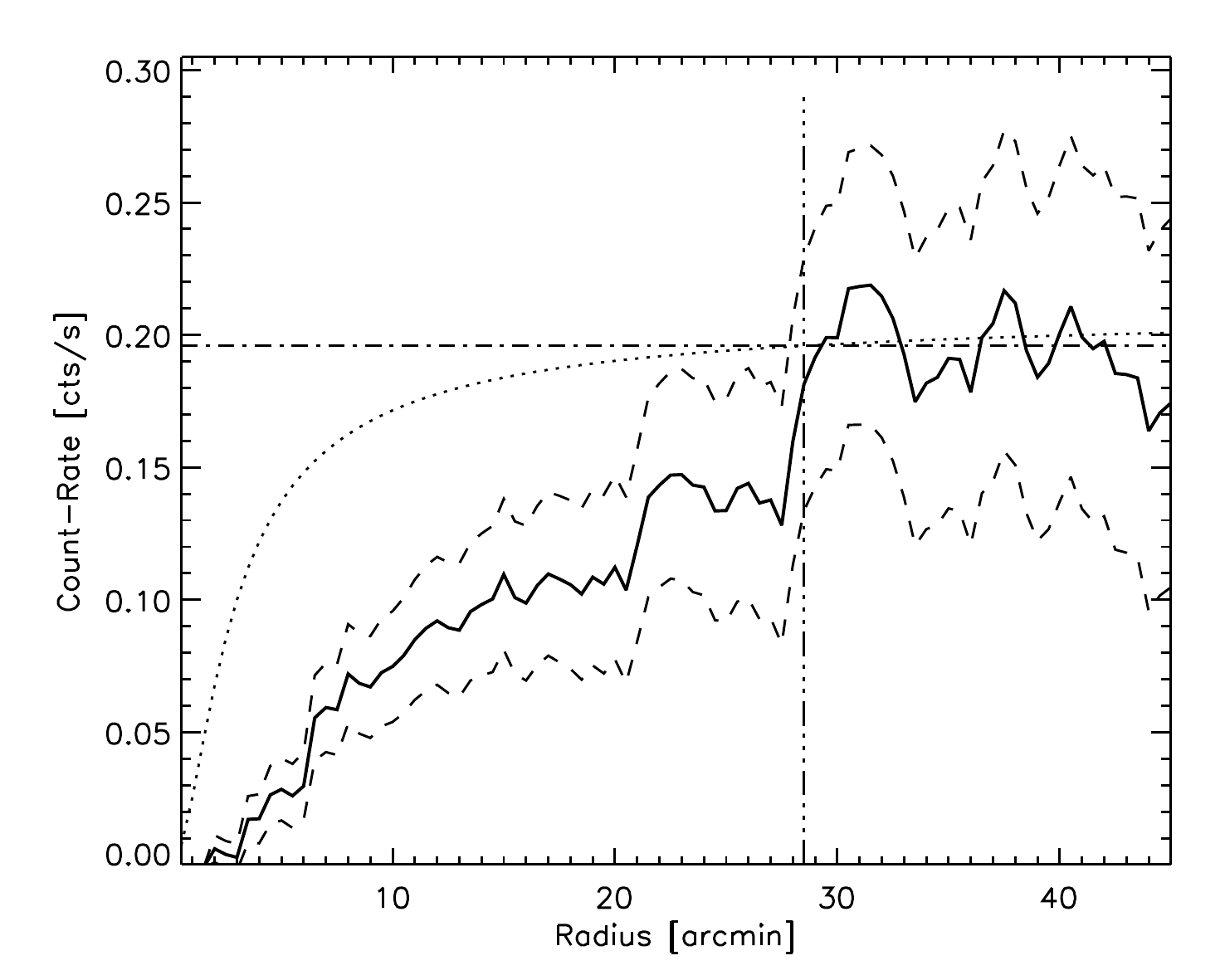}
\includegraphics[width=0.27\hsize]{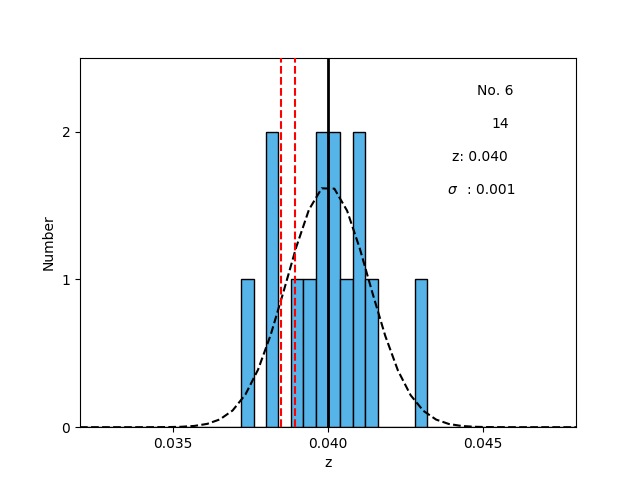}
\caption{\footnotesize{Group no.~$6$}} \label{fig:1f}
\end{subfigure}
\end{sidewaysfigure*}

\newpage

\begin{sidewaysfigure*}
\centering
\ContinuedFloat
\hspace*{\fill}
\begin{subfigure}{\textwidth}
\centering
\includegraphics[width=0.24\columnwidth]{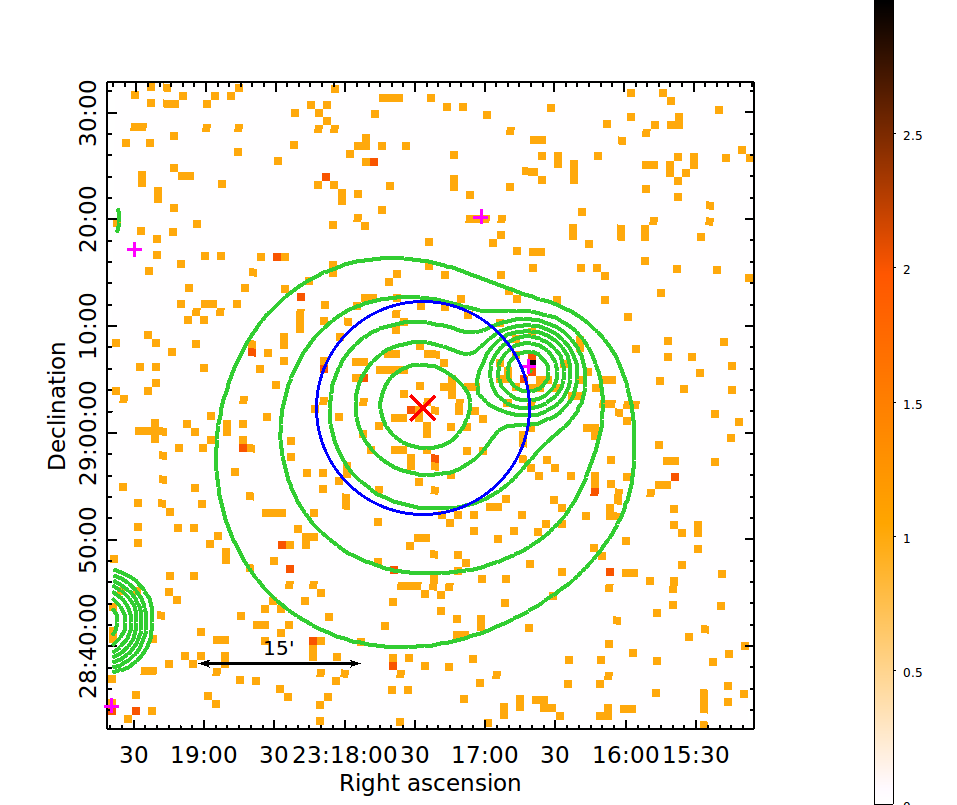}
\includegraphics[width=0.2\columnwidth]{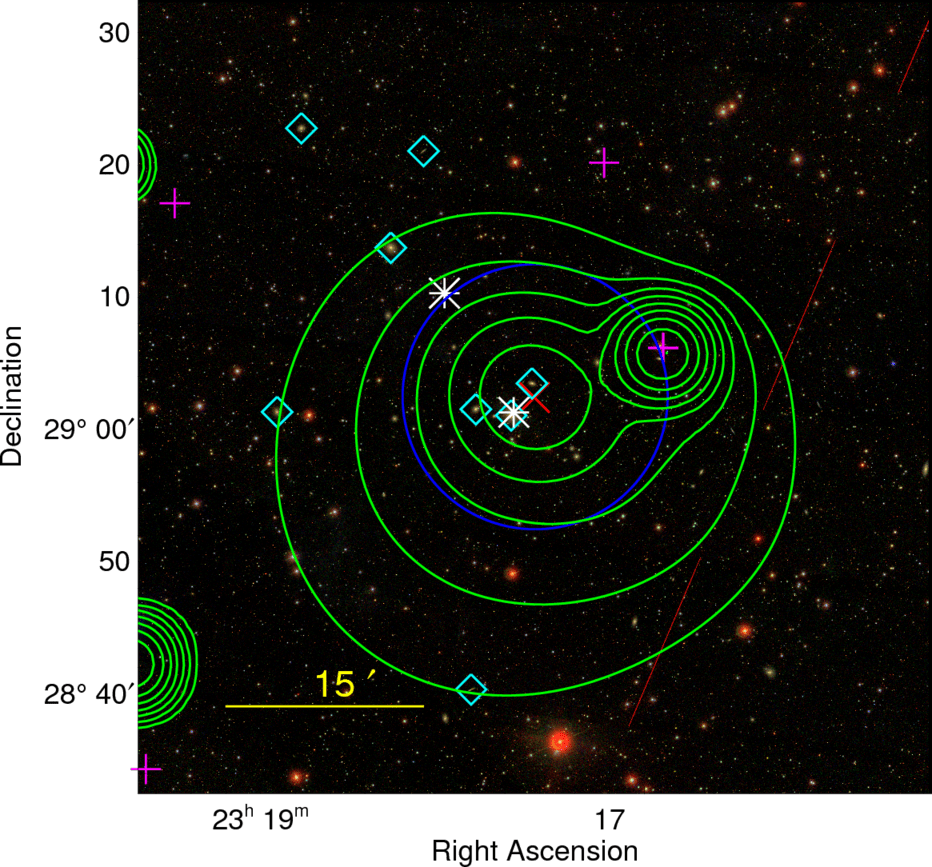}
\includegraphics[width=0.24\linewidth]{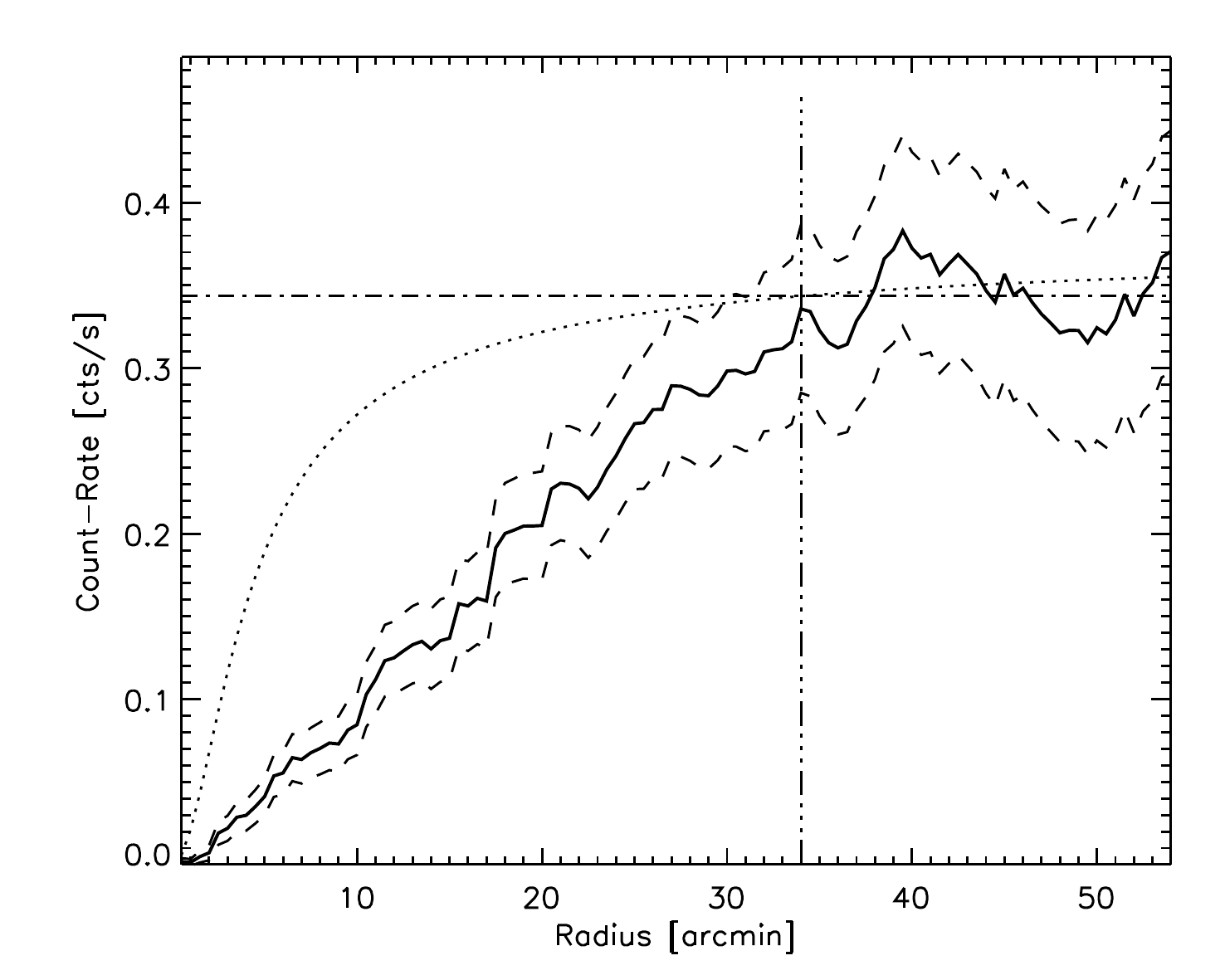}
\includegraphics[width=0.27\hsize]{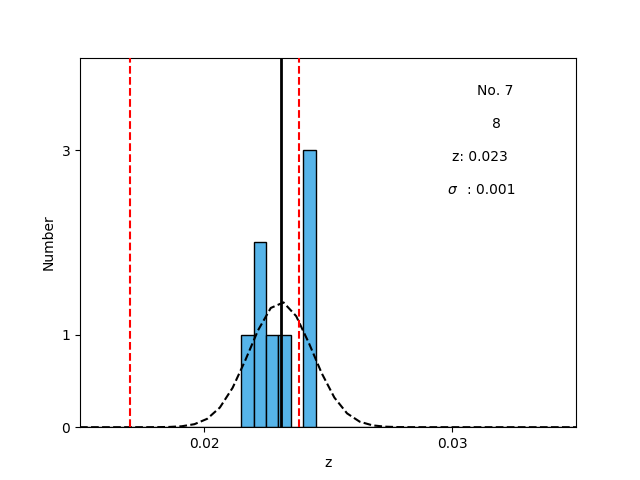}
\caption{\footnotesize{Group no.~$7$}} \label{fig:1g}
\end{subfigure}
\hspace*{\fill}
\begin{subfigure}{\textwidth}
\centering
\includegraphics[width=0.24\columnwidth]{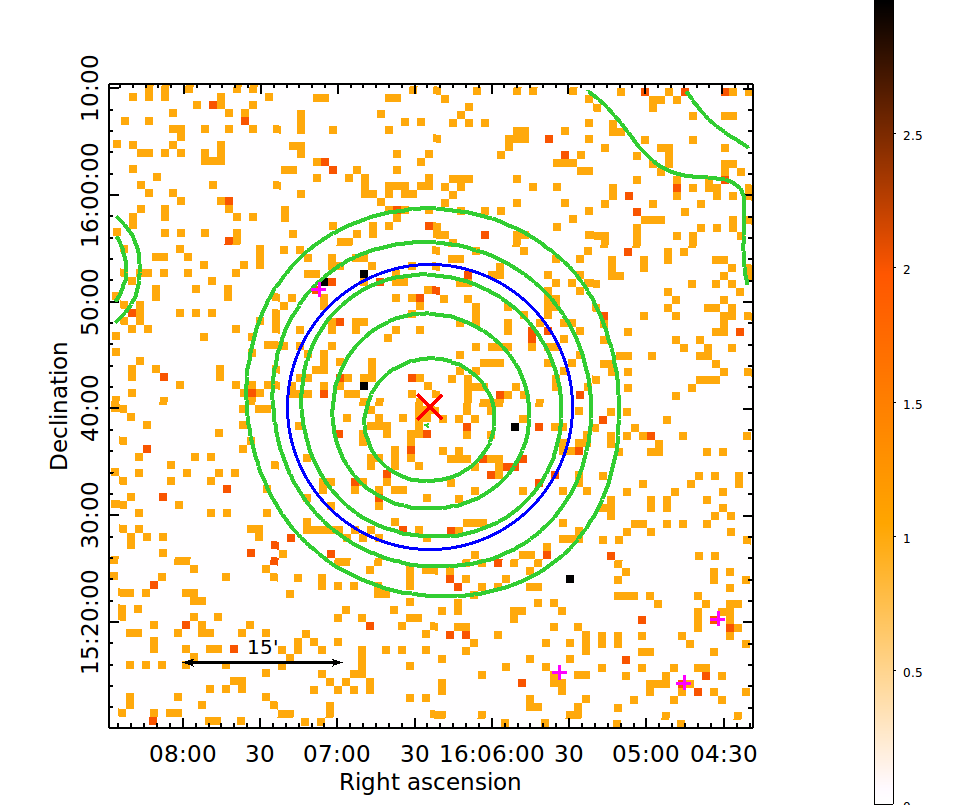}
\includegraphics[width=0.2\columnwidth]{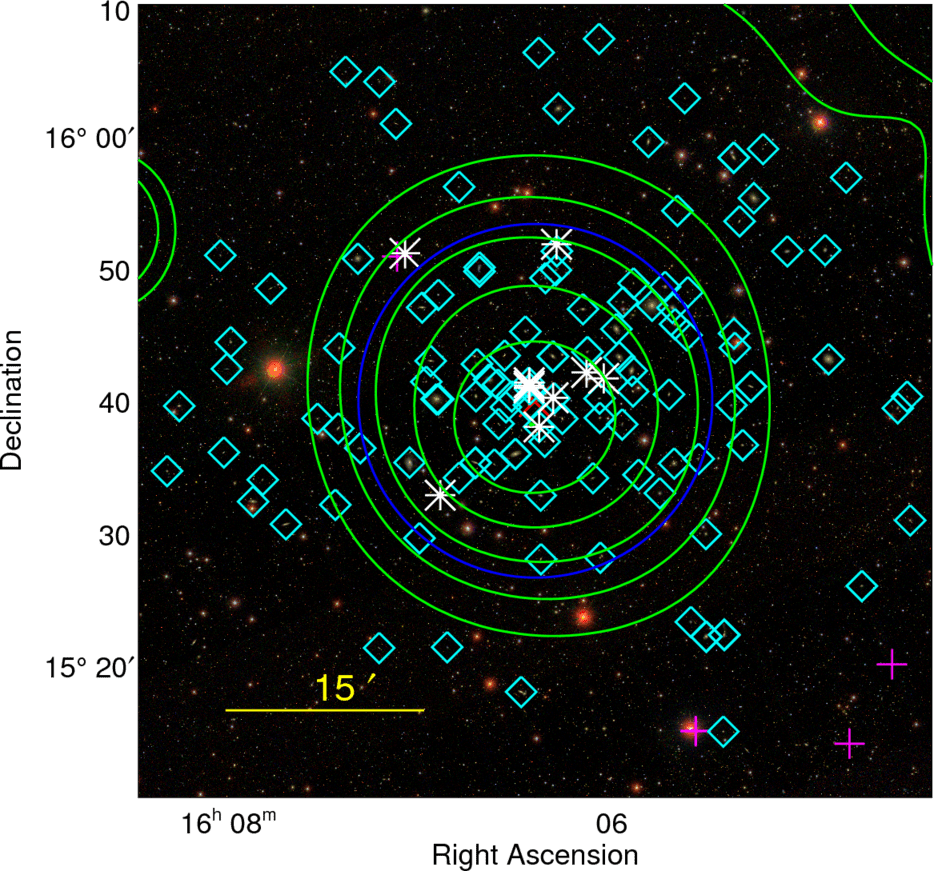}
\includegraphics[width=0.24\linewidth]{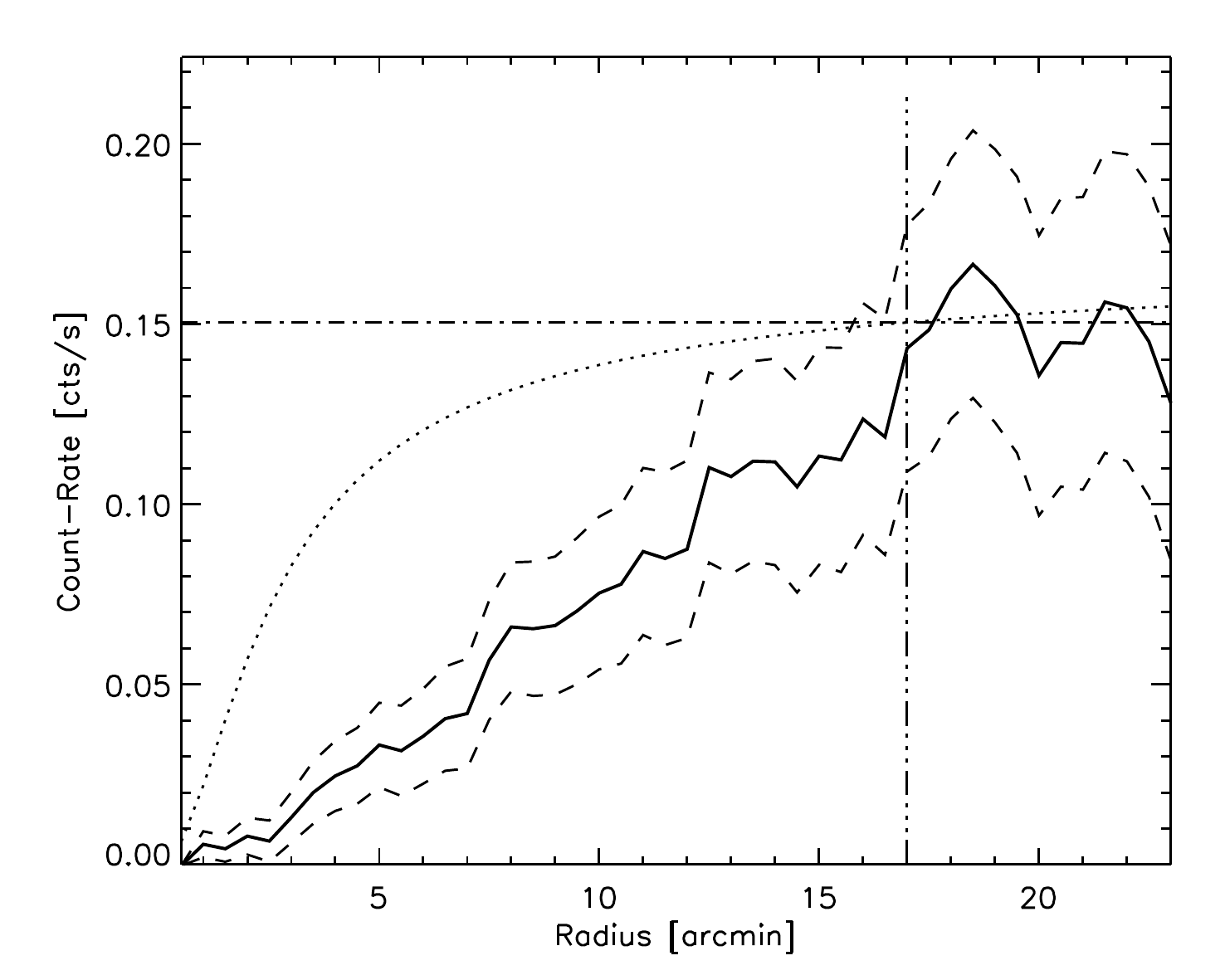}
\includegraphics[width=0.27\hsize]{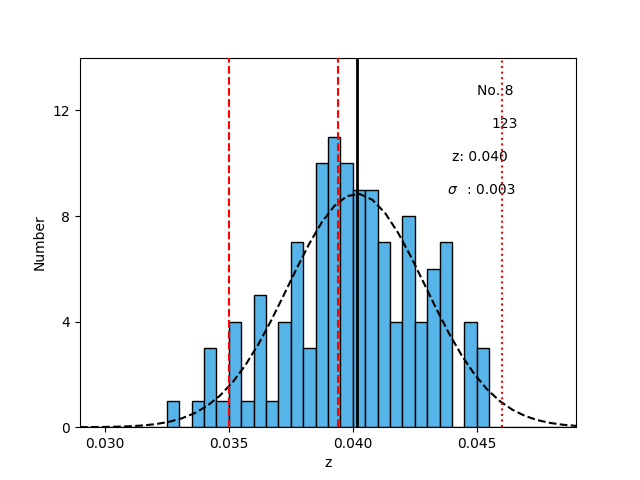}
\caption{\footnotesize{Group no.~$8$}} \label{fig:1h}
\end{subfigure}
\hspace*{\fill}
\begin{subfigure}{\textwidth}
\centering
\includegraphics[width=0.24\columnwidth]{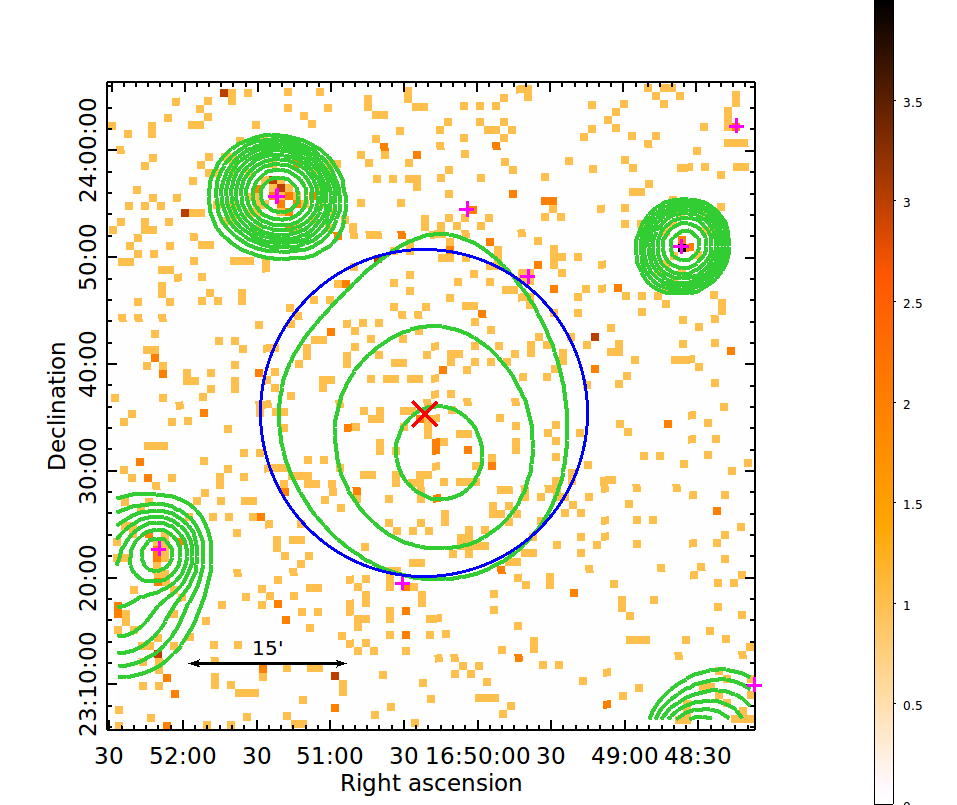}
\includegraphics[width=0.2\columnwidth]{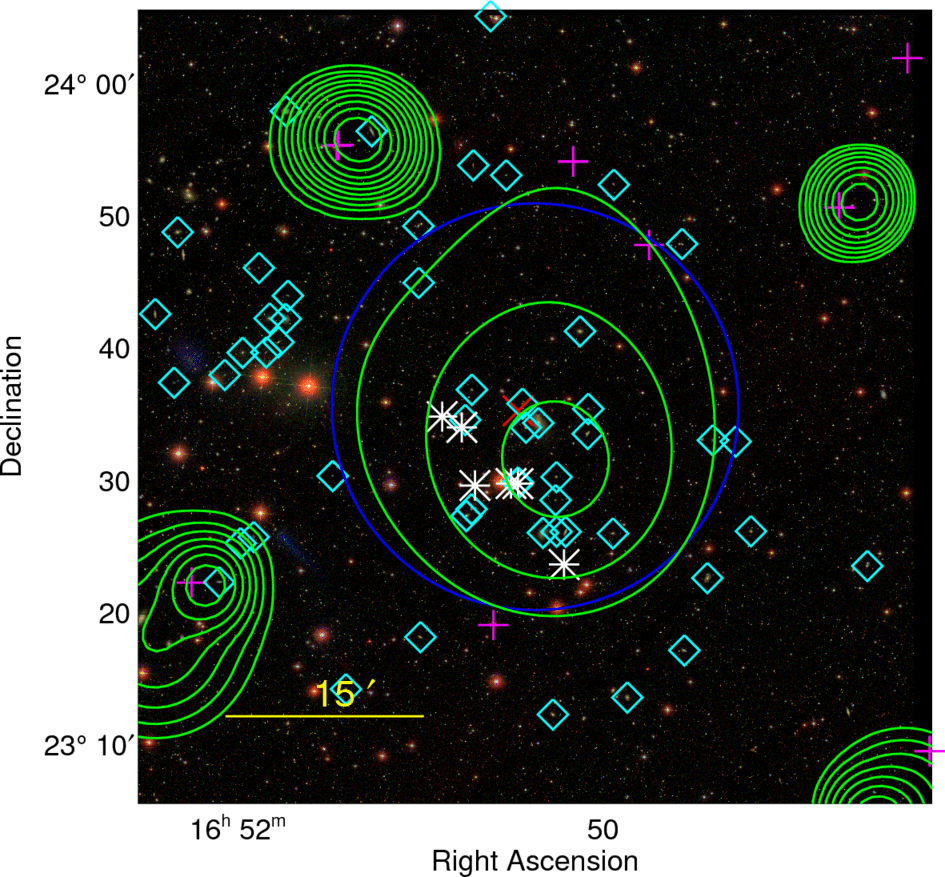}
\includegraphics[width=0.24\linewidth]{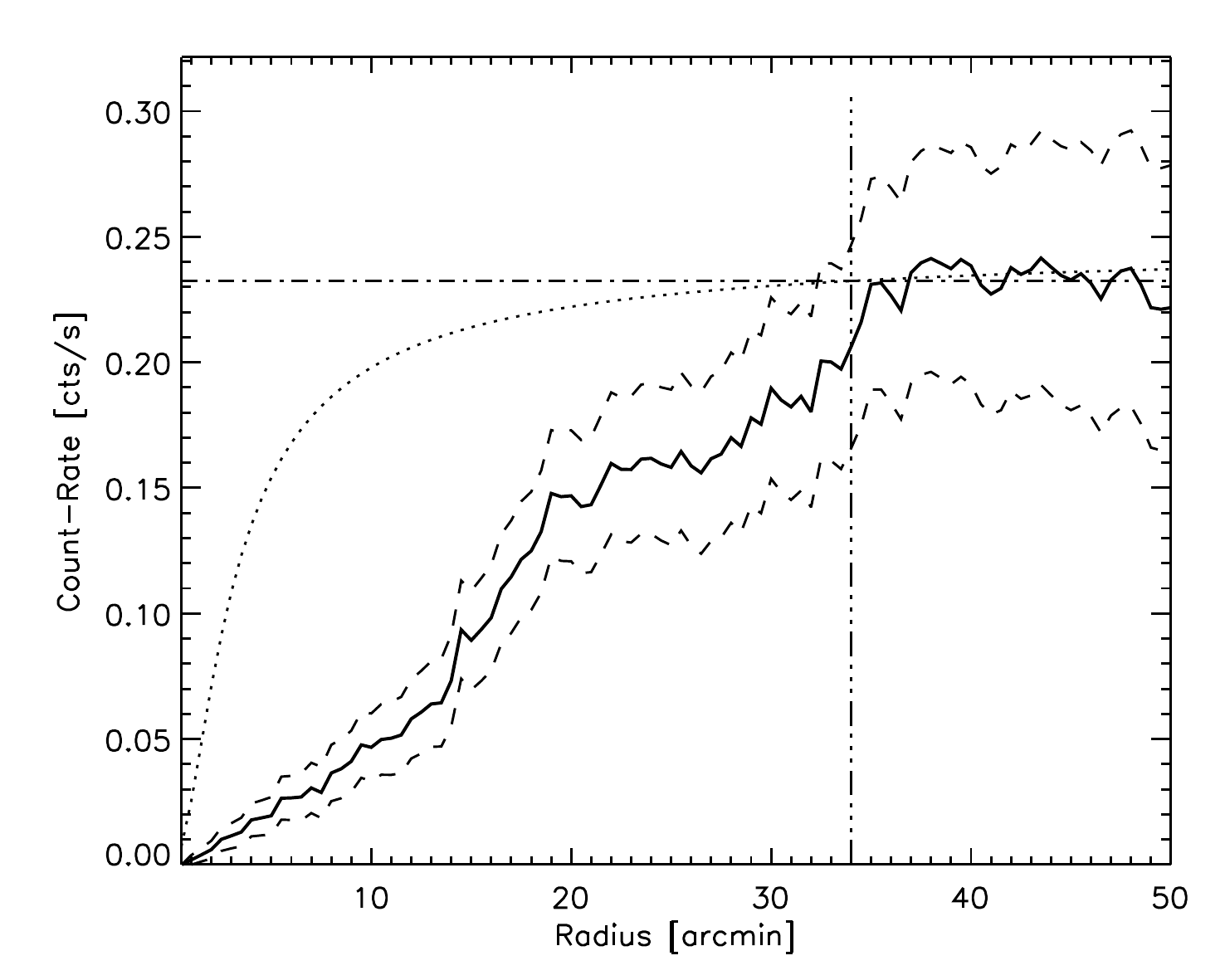}
\includegraphics[width=0.27\hsize]{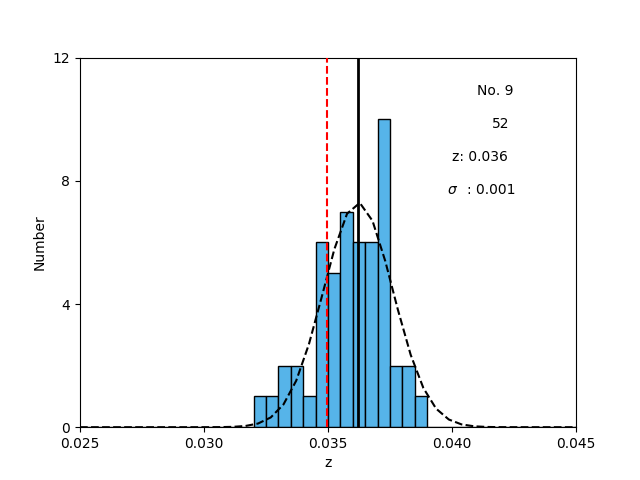}
\caption{\footnotesize{Group no.~$9$}} \label{fig:1i}
\end{subfigure}
\end{sidewaysfigure*}

\newpage

\begin{sidewaysfigure*}
\centering
\ContinuedFloat
\hspace*{\fill}
\begin{subfigure}{\textwidth}
\centering
\includegraphics[width=0.24\columnwidth]{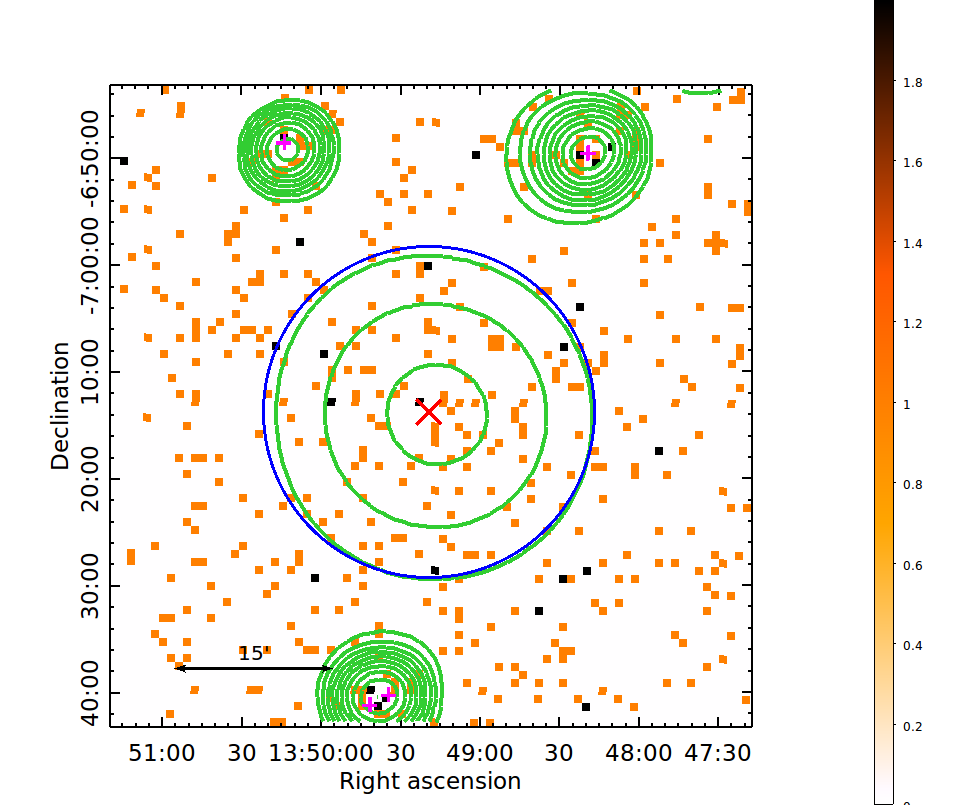}
\includegraphics[width=0.195\columnwidth]{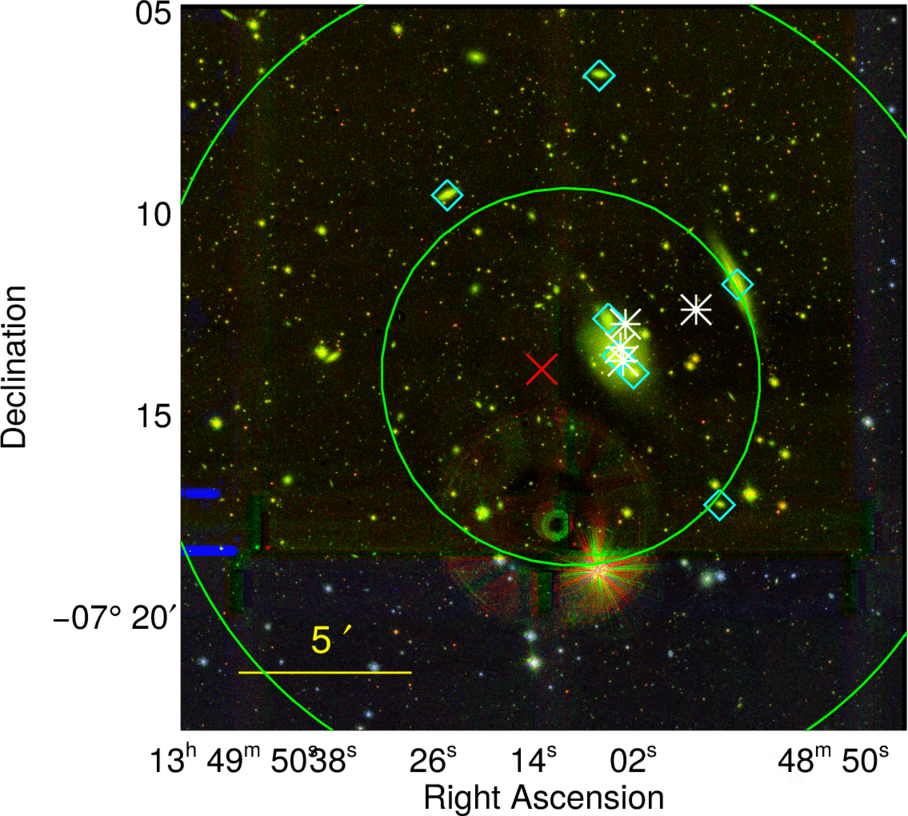}
\includegraphics[width=0.24\linewidth]{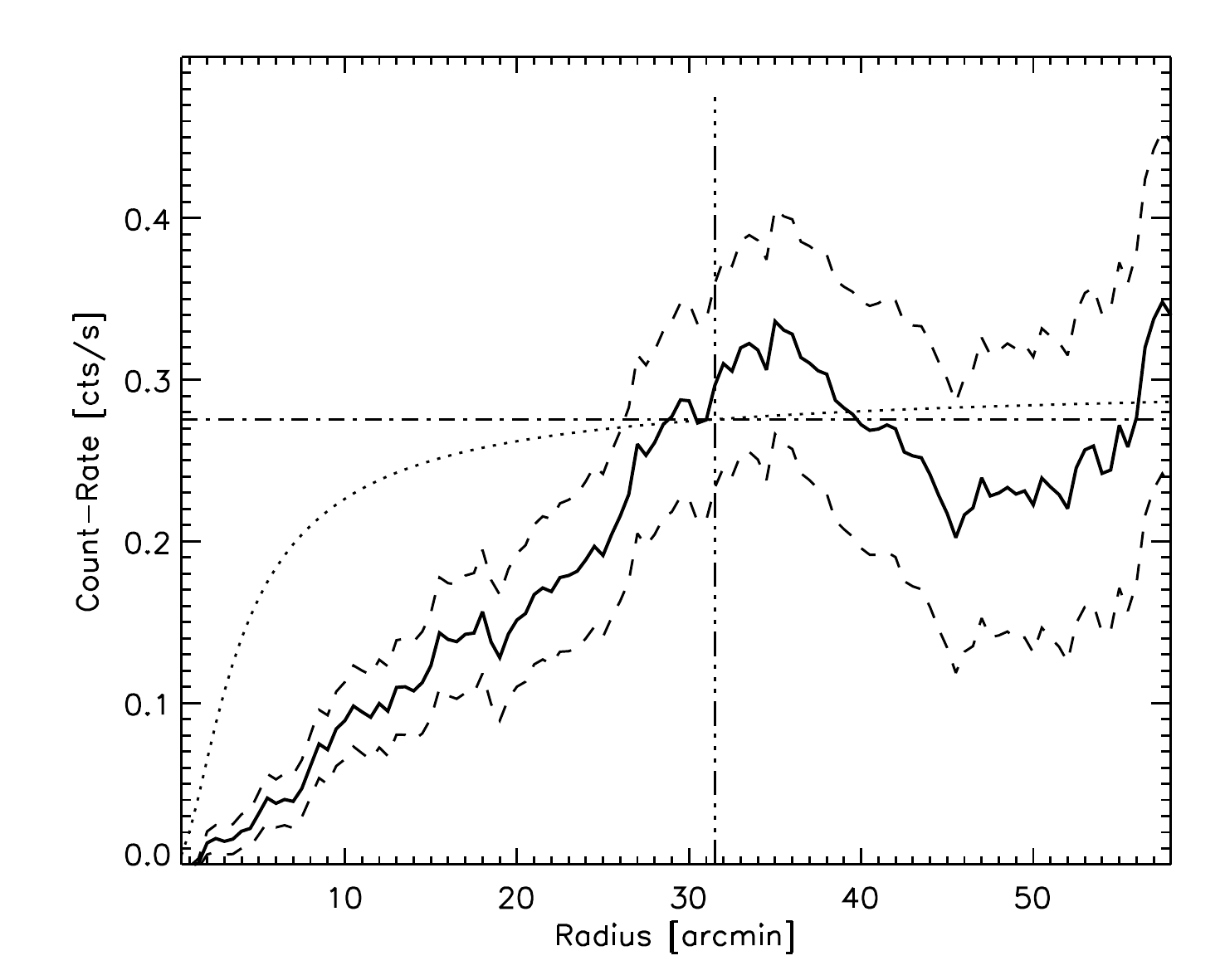}
\includegraphics[width=0.27\hsize]{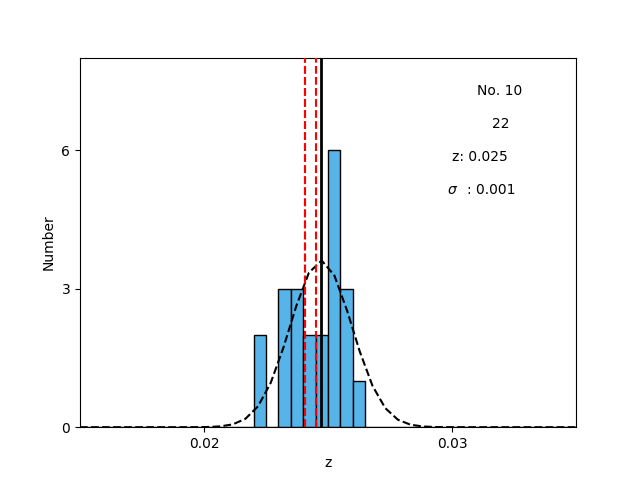}
\caption{\footnotesize{Group no.~$10$}} \label{fig:1j}
\end{subfigure}
\hspace*{\fill}
\begin{subfigure}{\textwidth}
\centering
\includegraphics[width=0.24\columnwidth]{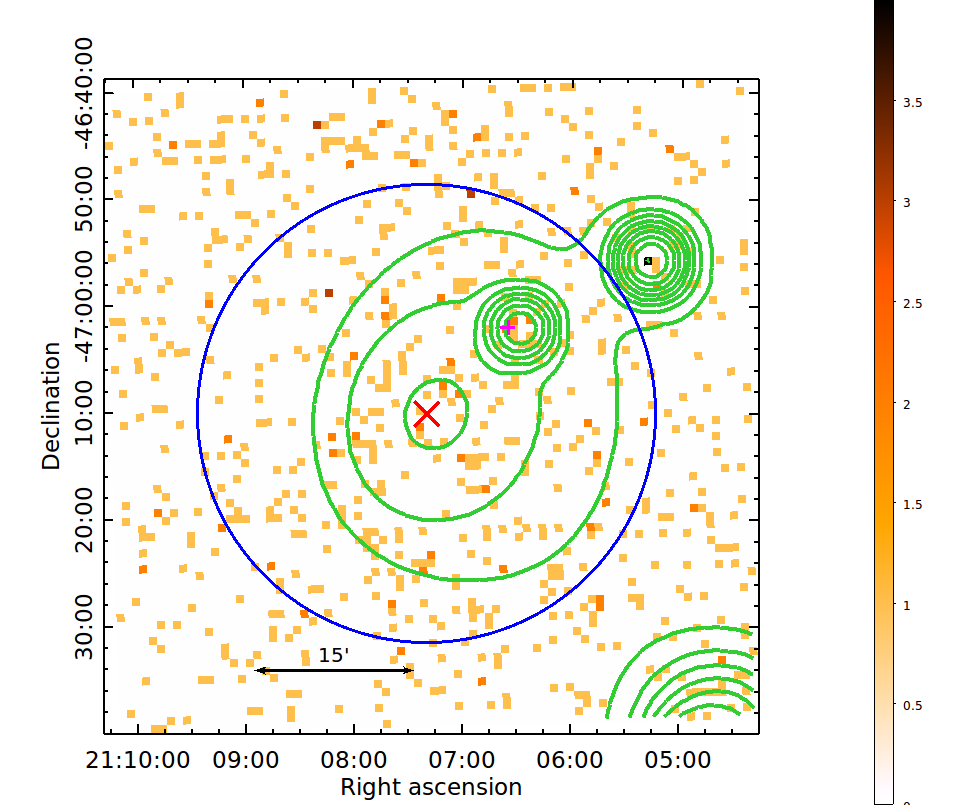}
\includegraphics[width=0.2\columnwidth]{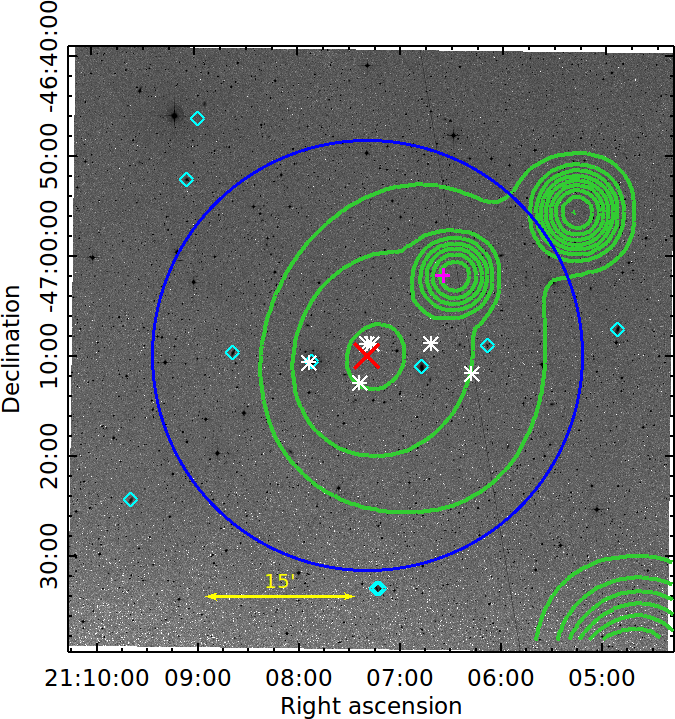}
\includegraphics[width=0.24\linewidth]{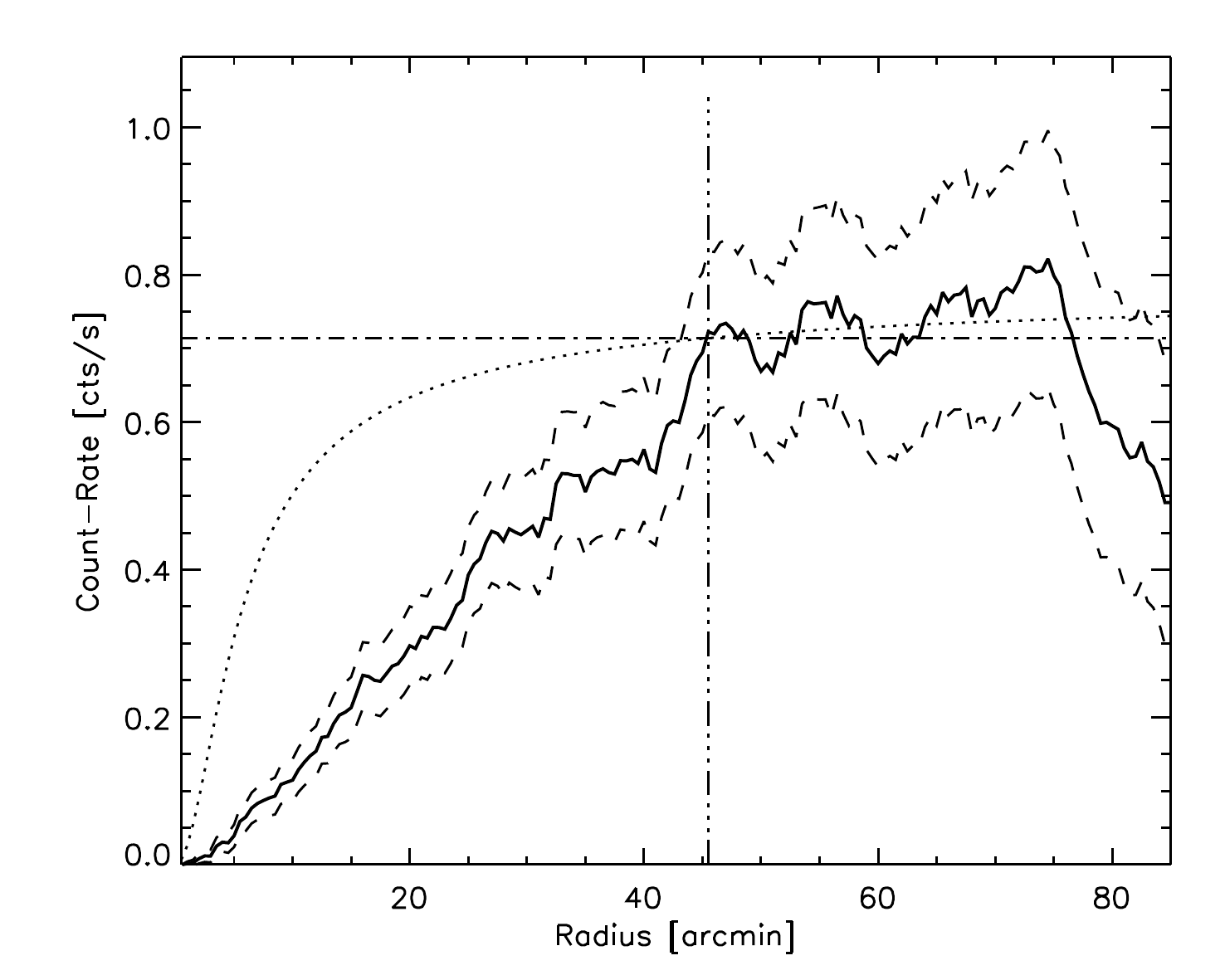}
\includegraphics[width=0.27\hsize]{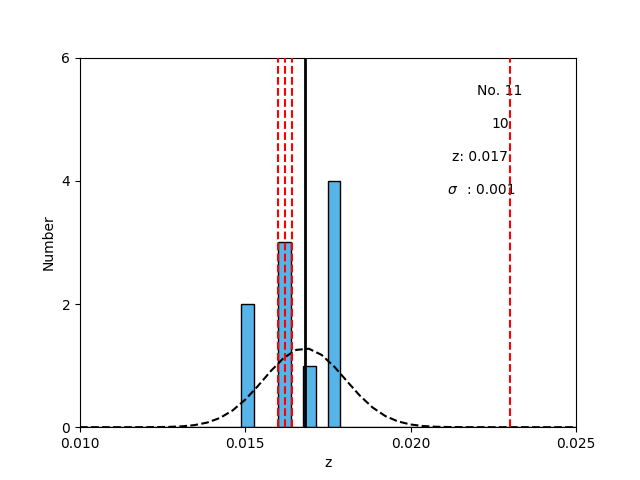}
\caption{\footnotesize{Group no.~$11$}} \label{fig:1k}
\end{subfigure}
\hspace*{\fill}
\begin{subfigure}{\textwidth}
\centering
\includegraphics[width=0.24\columnwidth]{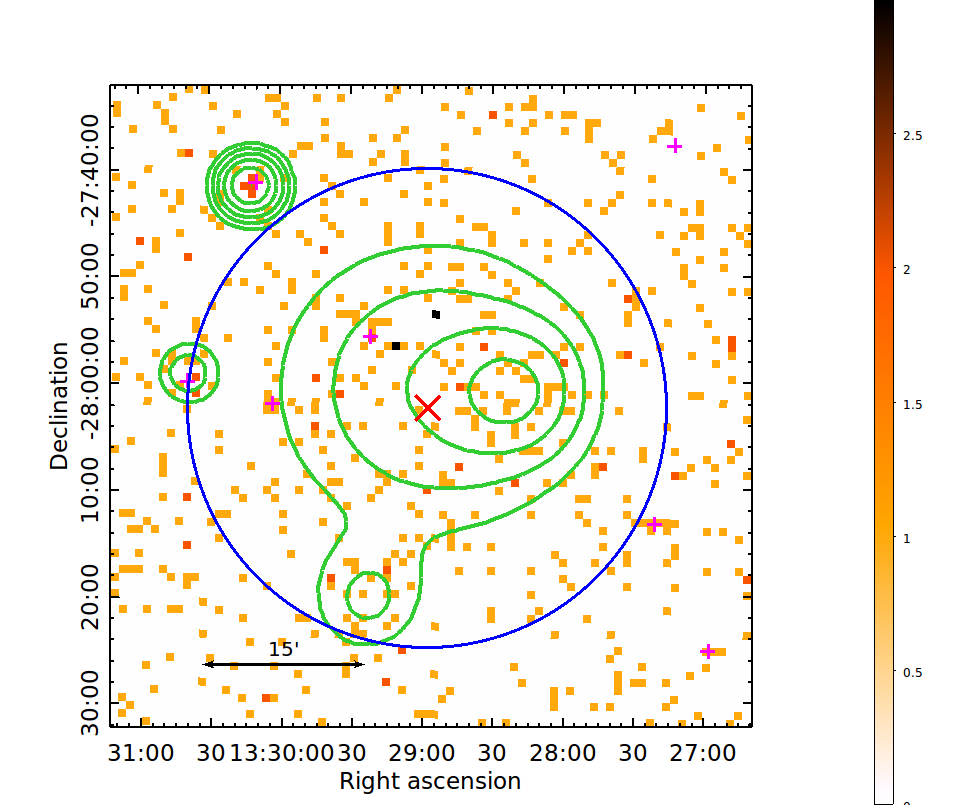}
\includegraphics[width=0.2\columnwidth]{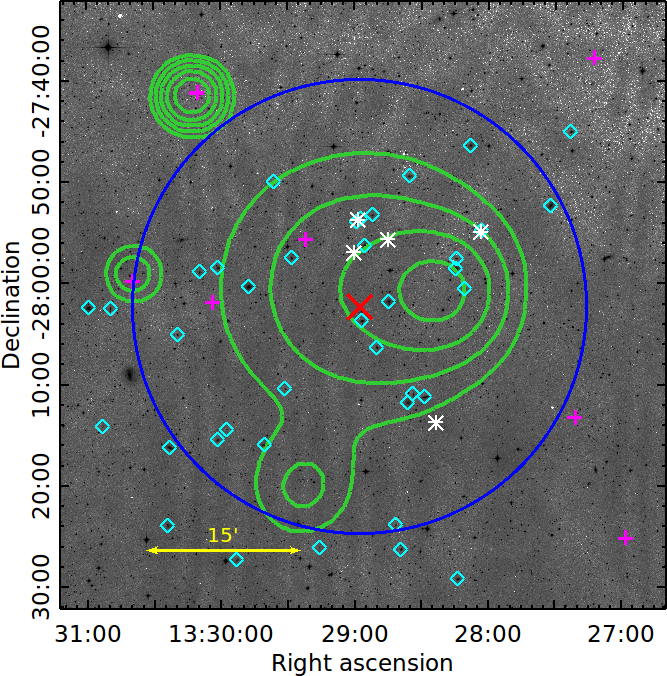}
\includegraphics[width=0.24\linewidth]{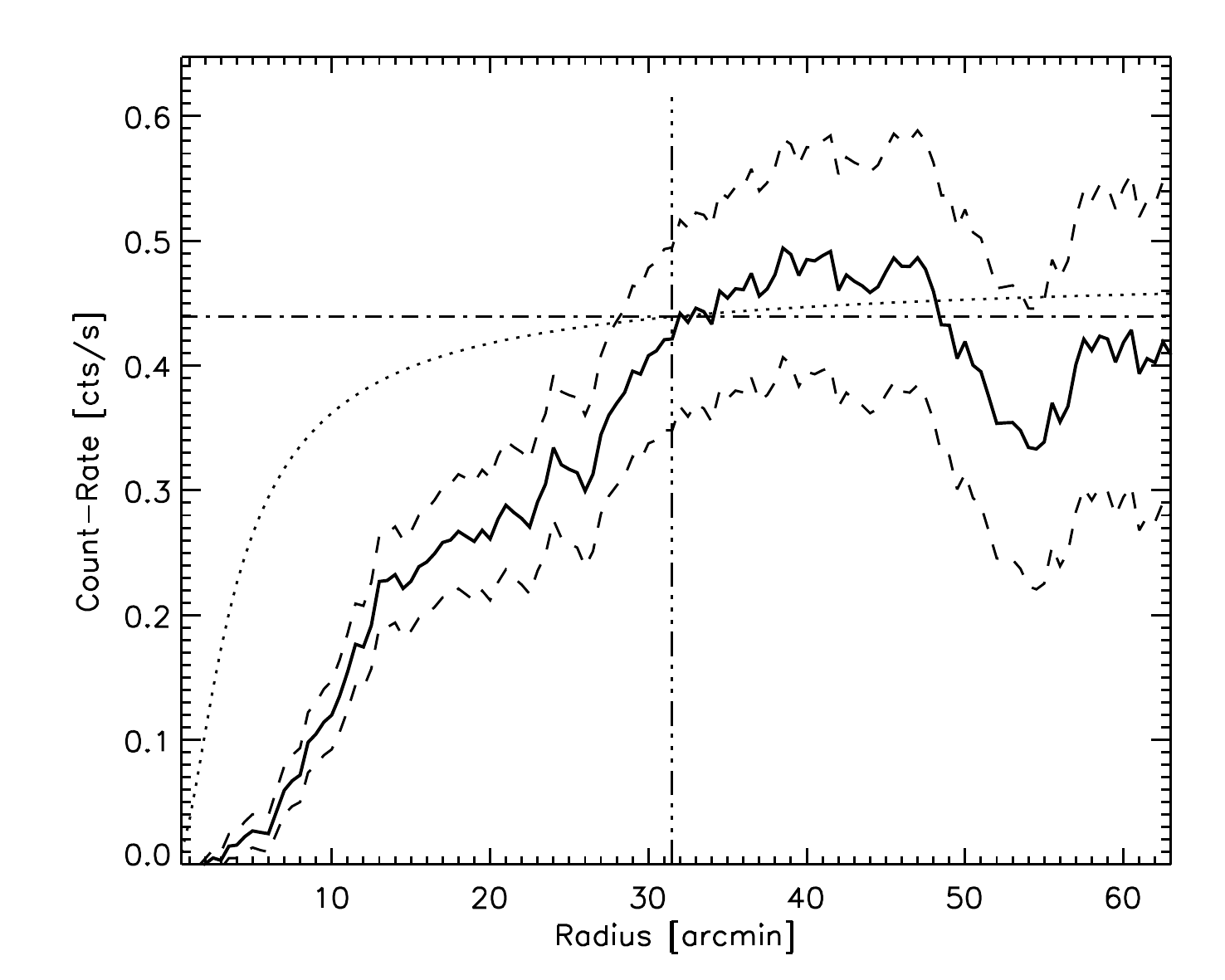}
\includegraphics[width=0.27\hsize]{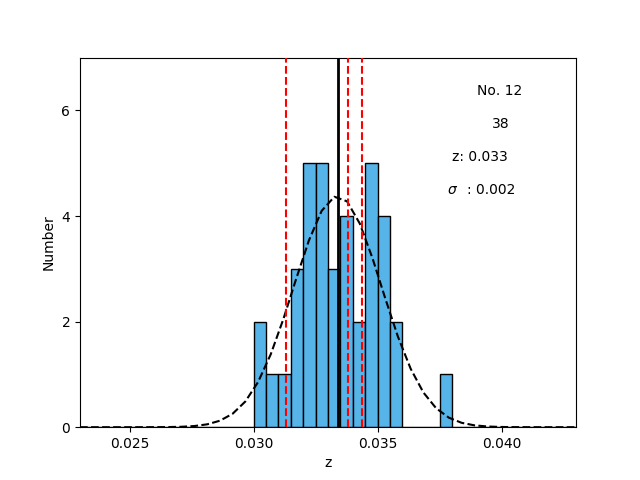}
\caption{\footnotesize{Group no.~$12$}} \label{fig:1l}
\end{subfigure}
\end{sidewaysfigure*}

\newpage

\begin{sidewaysfigure*}
\centering
\ContinuedFloat
\hspace*{\fill}
\begin{subfigure}{\textwidth}
\centering
\includegraphics[width=0.24\columnwidth]{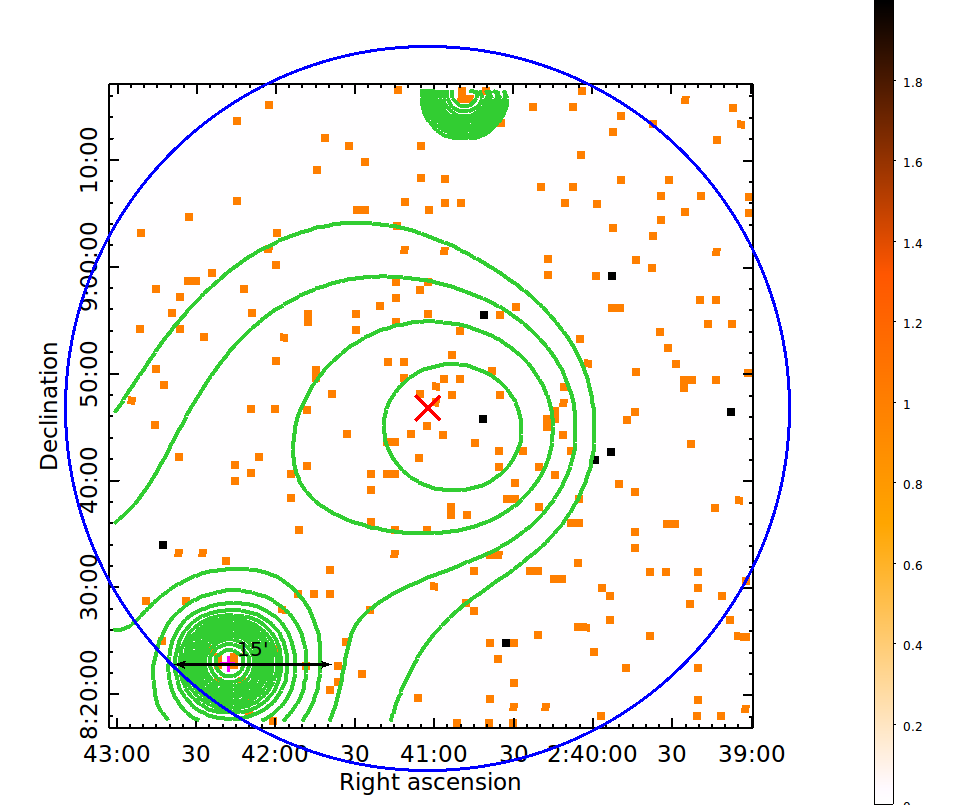}
\includegraphics[width=0.22\columnwidth]{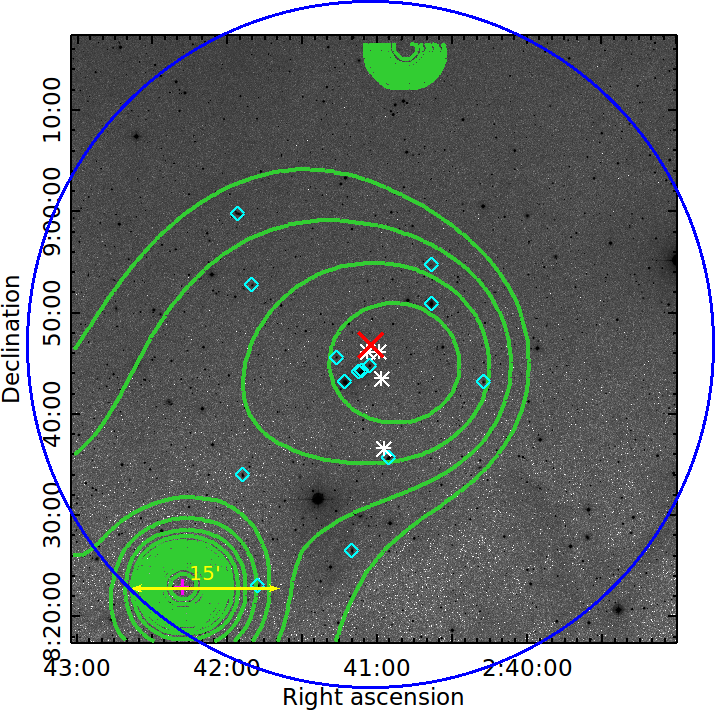}
\includegraphics[width=0.24\linewidth]{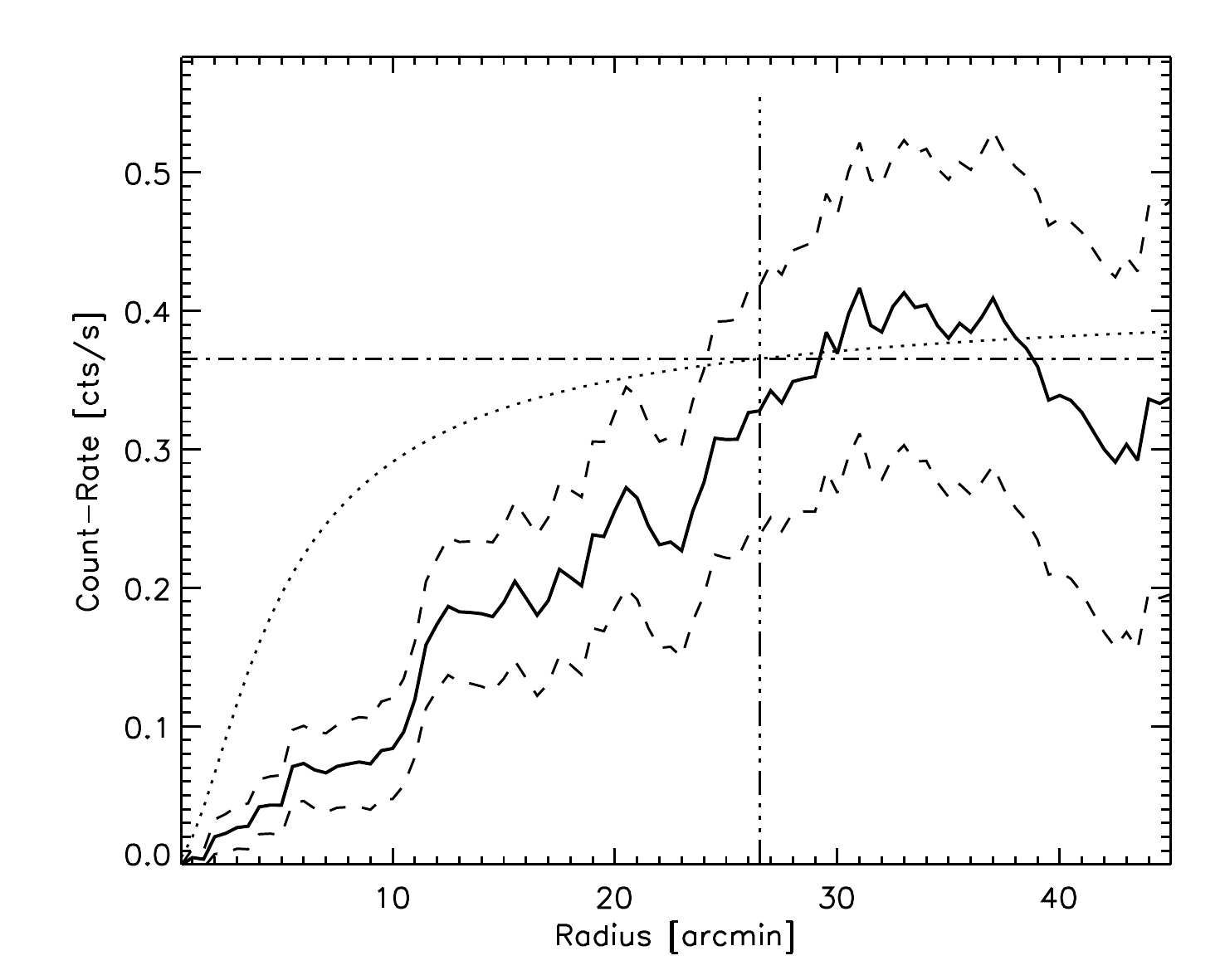}
\includegraphics[width=0.27\hsize]{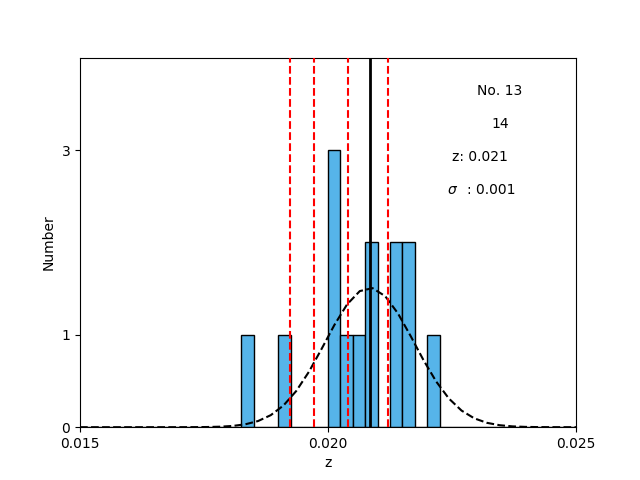}
\caption{\footnotesize{Group no.~$13$}} \label{fig:1m}
\end{subfigure}
\caption{\footnotesize{From {\it left} to {\it right}: RASS photon images ({\it left}), optical images, integrated count-rate profiles, and spectroscopic redshift histograms ({\it right}) for the galaxy groups in our pilot sample. The optical images are from ATLAS (composed of $g,~r,~i$ bands, group no.~$2,~6,~10$), SDSS (composed of $g,~r,~i$ bands, group no. $1,~3,~7,~8,~9$) or DSS ($r$~band, other groups). The images have a $1\times1$~deg$^2$ size, except for the ATLAS images, which are $0.30\times0.30$~deg$^2$. The red cross shows the position of the group candidate. The magenta plus signs display the location of 1RXS or 2RXS sources. The cyan diamonds show the positions of galaxies used for redshift determination. The white asterisks show the positions of known optical galaxy groups/clusters obtained from NED. The green contours are obtained from the corresponding wavelet-filtered images, and the blue circle shows the extent value (i.e., the core radius) of the group. The integrated count-rate profile is background subtracted (solid line). The dashed curves give the $1\sigma$ statistical error of the count-rate. The vertical and horizontal dot-dashed lines indicate the aperture radius and the count-rate used for the primary photometric results. The dotted curve shows the $\beta$-model of the cluster modelling used to extrapolate to $r_{500}$ (see Section~\ref{sec:flux_estim} for further details). In the spectroscopic redshift histograms the redshift values are taken for all galaxies within a box of $1\times1$~deg$^2$ centered on the group position with $z<0.10$ from NED. The total number of galaxies used in each redshift determination is given by the second number displayed in the upper right corner of each panel (see also Table~\ref{tab:candi}). The redshift distribution is fitted with a Gaussian function using $3\sigma$ clipping (black dashed curve). The final redshift of each group is taken from the peak of the fitting, and it is shown by the third number displayed in the upper right corner of each panel and by the black solid line. The red lines show the redshifts of identified groups or clusters within $15$~arcmin of the position of our candidates, red dashed line for spectroscopic redshift and the red dotted line for photometric redshift (see Section~\ref{sect:notes} for further details). }}
\label{fig:clustercandidates}
\end{sidewaysfigure*}

\end{document}